\documentclass[nofootinbib,aps,amsmath,twocolumn,superscriptaddress]{revtex4}
\usepackage{graphicx}
\usepackage{bm}
\usepackage{epsfig}

\newcommand{\lsim}{\mathrel{\rlap{\lower4pt\hbox{\hskip1pt$\sim$}}
    \raise1pt\hbox{$<$}}}         
\newcommand{\gsim}{\mathrel{\rlap{\lower4pt\hbox{\hskip1pt$\sim$}}
    \raise1pt\hbox{$>$}}}         

\newcommand{\be}{\begin{equation}}
\newcommand{\ee}{\end{equation}}
\newcommand{\bea}{\begin{eqnarray}}
\newcommand{\eea}{\end{eqnarray}}
\newcommand{\beq}{\begin{equation}}
\newcommand{\eeq}{\end{equation}}
\newcommand{\beqa}{\begin{eqnarray}}
\newcommand{\eeqa}{\end{eqnarray}}
\newcommand{\no}{\nonumber}

\begin{document}

 \title{On Indirect CP Violation and Implications for $D^0-\overline{D^0}$ and $B_s -\overline{B_s}$ mixing  }
  
 \author{Alexander L. Kagan and Michael D. Sokoloff} \address{Department of Physics, University
   of Cincinnati, Cincinnati, Ohio 45221, USA}
  
 \vskip .25in

 \begin{abstract}
The two kinds of indirect CP violation in neutral meson systems are related, in the absence of new weak phases in decay.  
The result is a model-independent expression relating CP violation in mixing, CP violation in the interference of 
decays with and without mixing, and the meson mass and width differences. It relates the semileptonic and time-dependent CP asymmetries;
and CP-conjugate pairs of time-dependent $D^0$ CP asymmetries.  CP violation in the interference of decays with and without mixing 
is related to the mixing parameters of relevance to model building: the off-diagonal mixing matrix elements $|M_{12} |$, $|\Gamma_{12} |$, and 
$\phi_{12} \equiv {\rm arg} (M_{12} /\Gamma_{12})$.  Incorporating this relation into a fit to the $D^0 -\overline{D^0}$ mixing data 
implies a level of sensitivity to $|\phi^D_{12} |$ of 0.10 [rad] at $1\sigma$.  The formalism is extended to include new weak phases 
in decay, and in $\Gamma_{12}$.  The phases are highly constrained by direct CP violation measurements. Consequently, 
the bounds on $|\phi^D_{12} | $ are not significantly altered, and the effects of new weak phases in decay could be difficult to observe 
at a high luminosity flavor factory ($D^0 $) or at the LHC ($B_s $) via violations of the above relations, unlike  
in direct CP violation. 
\end{abstract}

 \vskip .2in

 \maketitle
\section{Introduction}\label{sec:introduction}
There are two kinds of indirect CP violation in neutral meson decays,
CP violation in pure mixing (CPVMIX) and CP violation in the interference
of decays with and without mixing (CPVINT) (see, for example,~\cite{nirlectures,nirlectures2}).
Let $M^0$ and $\overline{M^0} $  be the interaction eigenstates of a neutral meson system. 
Indirect CP violation in pure mixing 
is due to a non-vanishing relative phase, $\phi_{12} = {\rm arg}(M_{12} /\Gamma_{12})$, between the dispersive ($M_{12}$) and absorptive ($\Gamma_{12}$)
parts of the $\overline{M^0} -  M^0$ transition amplitude.  It is responsible for 
CP asymmetries in semileptonic decays ($\overline{M^0}, M^0 \to \ell^\pm X$).  
Indirect CP violation in the interference of decays with and without mixing ($M^0 \to \overline{M^0} \to f$ and $M^0 \to f$)
can occur in decays to final states which are common to $M^0$ and $\overline{M^0}$, leading to 
time-dependent CP asymmetries.

Direct CP violation corresponds to different magnitudes for 
decay amplitudes related by CP conjugation.
It requires at least two amplitude contributions with different CP violating weak phases
and different CP conserving strong phases.
The weak phases present in the decay amplitudes in addition to the dominant Standard Model (SM) weak phase,
subsequently referred to as ``new weak phases in decay", can also lead to
unequal CPVINT measurements for different final states, and to
T-violating triple-product correlations for $VV$ final states \cite{valencia},
even if strong phase differences are absent.

In general, CPVINT
receives contributions from CPVMIX and from new weak phases in decay.
However, if the latter is absent, then CPVINT
originates solely from the mixing phase 
$\phi_{12}$, and therefore it must be connected to CPVMIX.
Consequently, two related formulae can be derived:
(i) an expression for CPVINT 
in terms of the mixing parameters $\phi_{12}$, $|M_{12}|$, and $|\Gamma_{12}|$, see Eq.~(\ref{eq:fiteqphi}).
Such a relation was first derived in the limit $M_{12} \ll \Gamma_{12}$ \cite{Bergmann:2000id};
(ii) a model-independent expression relating
 the four mixing observables, i.e., the two kinds of indirect CP violation
and the neutral meson mass and width differences, see Eq.~(\ref{eq:modindepreln}).
(i) allows a fit of the
three mixing parameters to the four observables to be performed.  
(ii) leads to model-independent correlations between 
time-dependent and semileptonic CP asymmetries.  It also leads to simple relations between CP-conjugate time-dependent CP asymmetries in
$D^0$ decays to non-CP eigenstates.

Examples in which the connection between the 
indirect CP asymmetries can be realized are provided
by the tree-level dominated decays, e.g.,
$K^0 \to \pi\pi$, $D^0 \to K^\pm \pi^\mp$, $D^0 \to K^+ K^- , \,\pi^+ \pi^-$, and $B_s \to J/\Psi \phi$.
Contributions to these decays beyond the SM tree-level charged current interactions
can be neglected in the SM itself, as well as
in many of its proposed extensions.  
Thus, the underlying hypothesis of no new weak phases in decay
is often valid.  
Of particular interest are applications to the $D^0$ and $B_s $ systems, where
non-vanishing indirect CP asymmetries would constitute a clear signal for new physics.
In many SM extensions they could be present at levels which can be measured at
ongoing, imminent, or planned experiments.


We review the neutral meson mixing and CP violation formalism in Sections II and III.
Attention is paid to the independence of physical observables with respect to the sign convention for the neutral meson mass or width differences.
In Section IV we derive 
the expression for CPVINT in terms of $\phi_{12}$, $|M_{12}|$, and $|\Gamma_{12}|$; the model-independent relation between CPVMIX and CPVINT;
and the resulting correlations 
among the time-dependent and semileptonic CP asymmetries.  
In the case of the $D^0$, we discuss singly Cabibbo suppressed (SCS), Cabibbo favored (CF), and doubly Cabibbo suppressed decays
to CP and non-CP eigenstates.  In the case of the $B_s$, we focus on $b \to c\bar c s$ mediated transitions, e.g., $B_s \to J/\Psi \phi$. 
In Section V a fit to the $D^0  -\bar D^0$ mixing data is carried out to determine the allowed ranges for 
$\phi_{12}^D$, $|M_{12}^D |$, and $| \Gamma_{12}^D  |$
in models with negligible new weak phases in the tree-level dominated decays.

In Section VI we discuss in detail how the above results would be modified by the appearance of subleading 
weak phases in the decay amplitudes, and in $\Gamma_{12}$.  
Order of magnitude bounds on these weak phases, hence on violations of 
the relations between CPVMIX and CPVINT, can be obtained from existing direct CP violation measurements.
It then follows that (i) the bounds on $\phi_{12}^D$ (and $|M_{12}^D |$, $| \Gamma_{12}^D  |$) do not change significantly, and (ii) it could be difficult to detect violations at currently allowed levels in 
the $D^0$ and $B_s$ systems, at a super $B$ factory and at the LHC, respectively. In fact, the existence of new weak phases in decay would be much easier to discover directly, via direct CP asymmetry measurements.
However, with sufficient statistics
it could be possible to isolate and measure shifts in ${\rm arg}(\Gamma_{12})$ (due to new physics, or to subleading 
$O(V^*_{us} V_{ub} / V^*_{cs} V_{cb} )$ SM contributions 
in $B_s$ mixing) from such violations.  We conclude in Section VII.

While this work was in progress, Ref.~\cite{Grossman:2009mn} appeared, which also explores the 
relation between the two indirect CP asymmetries, in the absence of new weak phases in decay.  
Our starting point for the derivation of a model-independent relation differs, in that it explicitly removes a discrete ambiguity in $\phi_{12} \leftrightarrow - \phi_{12}$, and allows us to obtain a simple general expression.
The reader is referred to~\cite{Grossman:2009mn} for a discussion of 
all four neutral meson systems.
Also see~\cite{Bigi:2009df} 
for a discussion, based on~\cite{Grossman:2009mn}, of correlations between 
time-dependent and semileptonic CP asymmetries in decays to CP eigenstates. 
After completion of this work, we discovered that our model-independent relation, Eq. (\ref{eq:modindepreln}), 
can be found in~\cite{Ciuchini:2007cw}.
We augment their presentation by providing its derivation from the neutral meson
mixing formalism, and by discussing its significance for relating  CPVMIX and CPVINT.
Our fit procedure for $D^0  -\overline{D^0}$ mixing differs by removing
the discrete ambiguity in $\phi_{12} \to \phi_{12} + \pi$.



\section{Formalism}\label{sec:formalism}
We begin with a summary of the formalism for neutral meson mixing and decays~\cite{nirlectures,nirlectures2}.  
The neutral meson mass eigenstates are linear combinations of the strong interaction eigenstates $|M^0 \rangle$ and $|\overline{M^0} \rangle$,
\beqa
|M_{1,2} \rangle =  p | M^0 \rangle \pm q | \overline{M^0} \rangle\,,\label{eq:M1andM2}
\eeqa
where $|q|^2 + |p|^2 =1$.  
We define the mass and width differences as
\beqa x\equiv {m_2 - m_1 \over \Gamma },\ \ y \equiv {\Gamma_2 -\Gamma_1 \over 2 \Gamma}\,,\eeqa
where $\Gamma  \equiv (\Gamma_1 + \Gamma_2)/2$ is the average width.  

The decay amplitudes of the neutral mesons $M^0$ and $\overline{M^0}$ to CP conjugate final state $f$ and $\bar f$  
are denoted as
\beqa \label{eq:amplitudes}
A_f &=& \langle f | {\cal H} | M^0 \rangle\,, \ \ \overline A_f =  \langle f | {\cal H} | \overline{M^0} \rangle \,,\nonumber\\
A_{\overline f}& =& \langle \overline f | {\cal H} | M^0 \rangle\,, \ \ \overline A_{\overline f} =  \langle \overline f | {\cal H} | \overline{M^0} \rangle,
\eeqa
where ${\cal H}$ is the weak interaction effective Hamiltonian.
The decay amplitudes for the tree-level dominated decays 
can, in general, be written as 
\beqa\label{eq:fouramp}
A_f&=&A^T_{f} e^{+i\phi^T_{f}}[1+r_fe^{i(\delta_f+\phi_f)}],\nonumber\\
A_{\overline f}&=&A^T_{\overline f} e^{i(\Delta_{f}+\phi^T_{\overline f})} [1+r_{\overline f} e^{i(\delta_{\overline
f}+\phi_{\overline f})}] ,\nonumber\\
\overline{A}_{\overline f}&=&A^T_{f} 
e^{-i\phi^T_{f}}[1+r_fe^{i(\delta_f-\phi_f)}],\nonumber\\
\overline{A}_f&=&A^T_{\overline f} e^{i(\Delta_{f}-\phi^T_{\overline f})}
[1+r_{\overline f }e^{i(\delta_{\overline f}-\phi_{\overline f })}],
\eeqa
where $A^T_{f} $ and $A^T_{\overline f} $ are the magnitudes of the dominant SM tree-level contributions. The ratios $r_f$ and $r_{\overline f} $ are the relative magnitudes of subleading contributions containing new weak phases (they could arise from new physics, or from SM amplitudes with suppressed CKM structure).   
$\phi_f^T$, $\phi_{\overline f}^T$, $\phi_f$, and $\phi_{\overline f}$ are weak (CP violating) phases which appear with opposite signs in CP conjugate
amplitudes, and
$\Delta_f$, $\delta_f$,  and $\delta_{\overline f}$ are strong (CP conserving) phases which appear with the same signs in CP conjugate amplitudes.
All of the quantities entering Eq. (\ref{eq:fouramp}), except the weak phases, are understood to be phase space dependent for 3-body and higher final states.

In the case of decays to CP eigenstates, $\Delta_f = 0 (\pi)$ for 
CP even (odd) final states.
Eq. (\ref{eq:fouramp}) therefore reduces to
\beqa\label{eq:twoamp}
A_f&=&A^T_{f} e^{+i\phi^T_{f}}[1+r_fe^{i(\delta_f +  \phi_f)}],\nonumber\\
 \overline{A}_{f}&=& \eta_{f}^{CP} A^T_{f} e^{-i\phi^T_{f}} [1+r_{ f} e^{i(\delta_{f}-\phi_{ f})}] ,\eeqa
where $\eta_f^{CP} = + (-)$ for CP even (odd) final states.
The time-dependent CP asymmetries depend on
the universal quantity
\beqa
\lambda_f \equiv {q \over p }  {\overline A_f \over A_f } = -\eta_f^{CP} \left|{q\over p }\right| e^{i \phi}\,,
\label{eq:lambdaCP}\\\nonumber\eeqa
where $r_f$ of Eq. ({\ref{eq:twoamp}) is neglected in the equality, and $\phi$ is the relative weak phase between the mixing and decay amplitudes.  
Examples of decays to CP eigenstates include $D^0 \to K^+ K^-,\,\pi^+ \pi^-, K_s \pi^0$, and $B_s \to J/\Psi \phi ,\,D^{(*)+ }_s D^{(*)-}_s$.

In the ``pure-penguin" decay
$B_s \to \phi \phi$, $A_f^T $ is the magnitude of the Standard Model penguin amplitude. Neglecting 
$r_f$, 
the weak phase $\phi$ in Eq. (\ref{eq:lambdaCP}) is the same as in the tree-level $B_s$ examples above, up to a small correction 
$\delta \phi = 2\, {\rm  Im}(V^*_{us} V_{ub} /V^*_{cs} V_{cb})$.

For final states which are not CP eigenstates, the time-dependent CP asymmetries depend on
\beqa \lambda_f &\equiv& {q \over p }  {\overline A_f \over A_f } = - \left|{q\over p }\right| R_f e^{i (\phi + \Delta_f )}\,,\nonumber\\
\lambda_{\overline f} &\equiv &{q \over p }  {\overline A_{\overline f} \over A_{\overline f} } = - \left|{q\over p }\right| R_{\over f} e^{i (\phi -\Delta_f )}\,,
\label{eq:lambdanonCP}\eeqa
where $r_f$ and $r_{\overline f}$ of Eq. (\ref{eq:fouramp}) are neglected in the equalities, and
thus $R_{\overline  f}^{-1} = R_f \equiv A^T_{\overline f} / A^T_f  $.
Examples are $D^0 \to K^\mp  \pi^\pm $,$\,K^* K$ and $B_s \to D^\pm_s D^{*\mp}_s$.
The weak phase $\phi$ is the same in Eqs. ({\ref{eq:lambdaCP}) and ({\ref{eq:lambdanonCP}) 
for the $D^0$ decays (up to negligible corrections $\lsim |(V_{ub} V_{cb} )/ (V_{us} V_{cs})|\sim 10^{-3}$)
and tree-level $B_s$ decays listed above.

The $\overline{M^0} - M^0$ transition amplitudes are 
\beqa \langle   M^0 | H | \overline{M^0} \rangle &=& M_{12} - {i\over 2 } \Gamma_{12}\,,\nonumber\\
 \langle   \overline{M^0} | H | M^0 \rangle& =& M^*_{12} - {i\over 2 } \Gamma^*_{12}\,,\eeqa
where $H$ is the $2 \times 2 $ effective Hamiltonian governing neutral meson mixing.
We define the mixing parameters
\beqa \hspace{-0.2cm}x_{12} \equiv 2 |M_{12} |/\Gamma,~y_{12} \equiv |\Gamma_{12}|/\Gamma,~\phi_{12} \equiv {\rm arg}(M_{12}/\Gamma_{12})
.\eeqa
The notation $x_{12}$, $y_{12}$ is borrowed from~\cite{Grossman:2009mn}. 
$\phi_{12}$ is a CP violating weak phase which is responsible for CP violation in mixing ($|q/p| \ne 1$).
Solving the eigenvalue problem yields
\beqa    (x - i y)^2  &=& x_{12}^2 - y_{12}^2  - i 2 x_{12} y_{12} \cos\phi_{12} \,, ~~{\rm or} \nonumber \\
x^2 - y^2 &=& x_{12}^2  - y_{12}^2\,,\ \ x y = x_{12} y_{12} \cos\phi_{12} \,,
\label{eq:eigenvalues}\eeqa
and
\beqa \label{eq:qoverp}
{q\over p   } = {-\Gamma (x - i y)  \over 2 (M_{12} -{i\over 2} \Gamma_{12} ) }= {-2 (M_{12}^* -{1\over 2} \Gamma_{12}^* )\over \Gamma (x-i y)}\,.
\eeqa

The phase transformation $|M^0 \rangle \to e^{i \theta} |M^0 \rangle $,
$|\overline{M^0} \rangle \to e^{-i \theta} |\overline{M^0} \rangle $ has no physical effects, due to conservation of Strangeness, Charm, or Beauty number by the strong interactions.  Indeed, it is easily seen that $\phi_{12}$, $\lambda_f $, $\lambda_{\overline f} $, $x$, and $y$
(which are related to, or are themselves observables) 
are invariant under these phase redefinitions \cite{nirlectures}.
Furthermore, the mass eigenstates are rotated by a common phase factor.

One is free to identify $M_2$ or $M_1$ with either the short-lived meson ($M_S$) or
the heavier meson ($M_H $), by redefining $q \to -q$.  This is equivalent to choosing a sign-convention for $y$, which in turn fixes the sign of 
$x$ via ${\rm sign }(\cos\phi_{12})$, or vice-versa.
Note that changing the sign-convention for $y$ (or $x$) takes $\lambda_f \to  -\lambda_f$, or equivalently, $\phi \to \phi + \pi$.
However, the combinations $y \,\lambda_f $ and $x\, \lambda_f $, or $y \cos\phi , y \sin\phi$ and $x \cos\phi , x \sin\phi$
are sign-convention independent, which is seen explicitly from
Eq. (\ref{eq:qoverp}). Thus, they are candidates for being related to physical observables.

Examples of CP conserving observables are ${\rm sign}(y \cos\phi )$ and ${\rm sign}(x \cos\phi )$.  In the limit of small  or no CP violation, respectively:
(i) $M_S $ would be approximately or exactly CP-even if and only if ${\rm sign}(y \cos\phi ) =+1$, and 
(ii) $M_H $ would be approximately or exactly CP-even if and only if ${\rm sign}(x \cos\phi ) =+1$.
This is seen from Eqs.~(\ref{eq:M1andM2}),~(\ref{eq:amplitudes}), and ~(\ref{eq:lambdaCP}) by requiring that  
CP-even ($M_+$) and CP-odd ($M_-$) states decay into CP-even and CP-odd final states, respectively.
In fact, in the $D^0$ system in the limit of CP conservation, the observable $y_{\rm CP}$, defined 
in Eq.~(\ref{eq:yCPdef}),
is equivalent to  \cite{Link:2000cu}
\beqa
y_{\rm CP} = {\Gamma (D_+ ) - \Gamma (D_-)\over \Gamma (D_+ ) + \Gamma (D_- ) }\,.
\eeqa
The world average is \cite{HFAGcharm},
\beqa \label{eq:yCPexp} 
y_{CP}=(1.07\pm 0.26) \%\,.
\eeqa
Taking into account that $|q/p| \approx 1$ and $|\sin\phi | \ll1$, see Eq. (\ref{eq:yCPqovpsinphi}), one finds that
$y_{CP} \approx y \cos\phi$ to very good approximation \cite{Bergmann:2000id}, thus explicitly realizing 
(i) above.


An alternative choice employed by the PDG~\cite{PDG} and HFAG~\cite{HFAGcharm} collaborations for the $K^0$ and $D^0$ systems,
is to identify $M_2$ with the would-be CP-even state in the limit of no CP violation.
This amounts to choosing a convention for $\phi$, i.e., $\phi \approx 0$ rather than $\phi \approx \pi$.  Given that in both systems the approximately CP-even state is $M_S$,
this choice is equivalent to the sign-convention $y>0$.


If $M_2$ were identified with $M_S $  ($y>0$),
Eq. (\ref{eq:eigenvalues}) would give
\beqa x &&\hspace{-0.25cm}=  {\rm sign}(\cos\phi_{12} )\, \times\nonumber\\
&&\hspace{-0.45cm} \left(x_{12}^2   - y_{12}^2 +\sqrt{ (x_{12}^2 + y_{12}^2 )^2 -4 x_{12}^2 y_{12}^2  \sin^2\phi_{12} }\right)^{1\over2}\!,
~~~~~\label{eq:xfiteq}\\
\hspace{-1.5cm}y &&\hspace{-0.25cm}= \nonumber \\
&&\hspace{-0.45cm} \left(y_{12}^2   - x_{12}^2 +\sqrt{ (x_{12}^2 + y_{12}^2 )^2 -4 x_{12}^2 y_{12}^2  \sin^2\phi_{12} }\right)^{1\over 2}\!.~~~~~\label{eq:yfiteq}
\eeqa
If, instead, $M_2 $ were identified with $ M_H $ ($x>0$),
then the factor ${\rm sign}(\cos\phi_{12})$ would be moved to the equation for $y$, with appropriate modifications for 
the choices $y<0$ or $x<0$.
These equations relate the the neutral meson mass and width differences to the underlying mixing parameters
$x_{12}$, $y_{12}$, and $\phi_{12}$.

\section{The CP asymmetries}\label{sec:CPasymmetries}
CP violation in pure mixing corresponds to $|q/p|\ne 1$.  It can be measured via the ``wrong-sign" semileptonic CP asymmetry, 
\beqa\label{semileptonic}
a_{\rm SL} &\equiv& {\Gamma (M^0 (t) \to \ell^- X) -\Gamma (\overline{M^0} (t) \to \ell^+  X) \over 
       \Gamma (M^0 (t) \to \ell^-  X) +\Gamma (\overline{M^0} (t) \to \ell^+  X)  }\,,\nonumber \\  
       &=&(| q /p |^4 - 1 )/(| q / p |^4 + 1)\,.
\eeqa
In the limit $||q/p|-1 | \ll 1$, which holds to good approximation for all four meson systems, 
$a_{\rm SL} = 2 (|q/p|-1) $.

The $D^0$ time-dependent decay rates into a final state $f$ can be written as (see, for example, \cite{nirlectures2})
\beqa\label{eq:tddr}
\hspace{-0.3cm}\Gamma(D^0(t)\to f)&=&{1\over2} e^{-\tau}|A_f|^2
\Big\{(1+|\lambda_f|^2) \cosh(y\tau)\nonumber\\
\hspace{-0.3cm}&&+(1-|\lambda_f|^2) \cos(x\tau)+2 {\rm Re}(\lambda_f)\nonumber \\
\hspace{-0.3cm}&&\times \sinh(y\tau)
-2 {\rm Im}(\lambda_f) \sin(x\tau) \Big\},\\
\label{eq:tddr-2}
\hspace{-0.3cm}\Gamma(\overline{D^0}(t)\to f)&=&{1\over 2} e^{-\tau}|\overline{A}_f|^2
\Big\{(1+|\lambda_f^{-1}|^{2}) \cosh(y\tau)\nonumber\\
\hspace{-0.3cm}&&+(1-|\lambda_f^{-1}|^{2}) \cos(x\tau)+2 {\rm Re}(\lambda_f^{-1})  \nonumber \\
\hspace{-0.3cm}&&\times \sinh(y\tau)
-2 {\rm Im}(\lambda_f^{-1}) \sin(x\tau)\Big\},
\eeqa
where $\tau \equiv  \Gamma_D  t$.

For $D^0$ decays to CP eigenstates the above expressions yield, to good approximation, purely exponential forms
due to the small values of $x$ and $y$,
\beqa \label{time-dep}
\Gamma({D}^0(t)\to f)&\propto&\exp[-\hat\Gamma_{D^0\to f}\ t],\no\\
\Gamma(\overline{D^0}(t)\to f)&\propto&\exp[-\hat\Gamma_{\overline{D^0}\to
  f}\ t].
\eeqa
The decay rate parameters are \cite{Bergmann:2000id}
\beqa \label{eq:Gammahat}
\hat\Gamma_{D^0\to f}&=&
\Gamma_D[1+\eta^{\rm CP}_f\,  \left|{q/ p}\right| (y\cos\phi-x\sin\phi)],\no\\
\hat\Gamma_{\overline{D^0}\to f}&=&
\Gamma_D[1+\eta^{\rm CP}_f\,\left|{p/ q}\right|(y\cos\phi+x\sin\phi)],
\eeqa
where $\phi$ is defined in Eq.~(\ref{eq:lambdaCP}), and $r_f$ has been neglected. 
Note that Eq. (\ref{eq:Gammahat}) applies to singly Cabibbo suppressed (SCS) 2-body decays (e.g.,
$D^0 \to K^+ K^-\,,\pi^+\pi^-$), and to 2-body decays in which both Cabibbo favored (CF) and doubly Cabibbo suppressed (DCS)
amplitudes contribute (e.g., $D^0 \to K_s \pi^0$).
(In the case of decays to CP eigenstates which are resonances or multi-body states, Eq. (\ref{eq:Gammahat}) is valid when ignoring
the interference of these amplitudes with other amplitudes in phase space, see below.)
One defines the CP violating combination (or lifetime CP asymmetry),
\beqa\label{eq:defdy}
\Delta Y_f\equiv\frac{\hat\Gamma_{\overline{D^0}\to f}-\hat\Gamma_{D^0\to f}}
{2\Gamma_D}=a^m+a^i,
\eeqa
where
\beqa\label{eq:amai}
a^m&=&-\eta^{\rm CP}_f\,{y\over 2} \cos\phi \, \left( \left|{q \over p}\right|-\left|{p\over  q}\right|\right) \,,\nonumber\\
a^i &=& \eta^{\rm CP}_f\,{x\over 2} \sin\phi \, \left(\left|{q\over p}\right|+\left|{p \over q}\right|\right) \,.
\eeqa
$ a^m $ and $a^i$ are the contributions due to CPVMIX ($|q/p|\ne 1$) and CPVINT ($\sin\phi \ne 0$), respectively, and are universal quantities.   
Note that they are independent of sign convention for $x$ or $y$.
Subleading, non-universal
corrections to $ \hat\Gamma_{D^0\to f}$, $\hat\Gamma_{\overline{D^0}\to {f}}$ due to $r_f \ne 0$ are discussed in Section VI.

In SCS $D^0$ decays to non-CP eigenstates (e.g., $D^0 \to K^* K$), the final states are essentially resonances or multi-body states.
The time-dependence of the decays is again exponential, to good approximation, and is independent of phase space if the interference 
of these amplitudes with other amplitudes is ignored.  
In general, in decays to resonances, or multi-body decays, 
the exponential decay rate parameters depend on phase space (e.g., for 3-body decays, the location in the Dalitz plot)
and give two CP violating combinations~\cite{Grossman:2006jg},
\beqa\label{eq:defdy-non}
\Delta Y_f&\equiv&\frac{\hat\Gamma_{\overline{D^0}\to
\overline f}-\hat\Gamma_{D^0\to f}}{2\Gamma_D}=a^m_f+a^i_f, \nonumber\\
\Delta Y_{\overline f}&\equiv&\frac{\hat\Gamma_{\overline{D^0}\to
f}-\hat\Gamma_{D^0\to \overline f}}
{2\Gamma_D}=a^m_{\overline f}+a^i_{\overline f},
\eeqa
where, neglecting $r_f$ and $r_{\overline f}$, 
\beqa\label{eq:amainonCP}
a^m_f  &=&-R_f\, {y'_f \over 2}  \cos\phi\, \left( \left|{q \over p}\right|-\left|{p\over q}\right|\right) \,,\nonumber\\
a^i_f &=&R_f\,{x'_f \over 2} \sin\phi\,\left(\left|{q\over  p}\right|+\left|{p\over q}\right|\right)\,,
\eeqa
(for $a_{\overline f}$, replace
$f \to {\overline f}$),
\beqa
\label{eq:xprimeyprime}
x^\prime_f&=&x\cos\Delta_f+y\sin\Delta_f,~~~
y^\prime_f=y\cos\Delta_f-x\sin\Delta_f,\no\\
x^{\prime}_{\overline f}&=&x\cos\Delta_f-y\sin\Delta_f,~~~
y^{\prime}_{\overline f}=y\cos\Delta_f+x\sin\Delta_f\,,\no\\
\eeqa
and $\phi$, $\Delta_f$ are defined in Eq. (\ref{eq:lambdanonCP}).
In SCS decays one expects $R_f = O(1)$,
implying that the CP asymmetries for non-CP eigenstates should be of same order as for
CP eigenstates.

The quantities $a^m_{f } $, $a^i_{f} $, $a^m_{\overline f } $, $a^i_{\overline f} $ 
are not universal for non-CP eigenstate final states, due to the presence of strong phases.
However, the latter can be determined, e.g., for 3-body decays, from Dalitz plot analyses.
For example, in the simple case of a single resonance, $K^* K$, in the Dalitz plot,
$\Delta_{K^* K}$ can be determined from the interference region
of $K^{+*} K^-$ with $K^{*- }K^+$ \cite{Rosner:2003yk}. 
Consequently, $x$, $y$, $|q/p|$, and $\phi$ can be determined (up to discrete ambiguities) in Dalitz plot analyses of final states such as 
$D^0 \to K_s\, K^\pm \pi^\mp$ and $D^0 \to \pi^- \pi^+ \pi^0$ \cite{Grossman:2006jg}.

In the case of CF and DCS decays to non-CP eigenstates, the time-dependence for $D^0$ decays to the ``wrong-sign" (WS) final states
$D^0 (t) \to \overline{f}$ and $\overline{D^0} (t) \to f$
is expanded to quadratic order in 
$\tau$ inside the curly brackets of Eqs. (\ref{eq:tddr}) and (\ref{eq:tddr-2}), due to the small values of $\tan^2\theta_c $, $x$, and $y$
 ($A_{\overline f}$ is chosen to be the DCS amplitude, e.g., $D^0 \to K^+ \pi^-$ or $\overline{f} = K^+ \pi^-$).
The result can, in general, be written as (we adopt a notation similar to the one used in the experimental analysis of $D^0 \to K^\mp \pi^\pm$ \cite{BelleKPi})
\beqa\label{eq:DKppim}
\Gamma[D^0 (t) &\to& \overline{f}]=e^{-\tau } |A_{f}|^2 \times\\
&& \hspace{-.45cm}\left[(R_f^{+})^{2}  +R_f^+ \,y^{\prime +}\,\tau + {(x^{\prime +})^2 +(y^{\prime +})^2 \over 4 } \,\tau^2 \right], 
\nonumber\\
\label{eq:DbarKmpip}
\Gamma[\overline{D^0} (t) &\to& f]=e^{-\tau} |\overline{A}_{\overline f}|^2 \times\\
&& \hspace{-.45cm}\left[(R_f^{-})^{2}  + R_f^- \,y^{\prime -}\,\tau + {(x^{\prime -})^2 +(y^{\prime -})^2 \over 4 } \,\tau^2 \right],
\nonumber\eeqa
where $R_f^+ = |A_{\overline f} /A_f|$ and $R_f^- = |\overline{A}_{f} /\overline{A}_{\overline f}|$ are the magnitudes
of the DCS to CF amplitude ratios for $D^0$ and $\overline{D^0}$ decays.
Neglecting $r_f$ and $r_{\overline f}$ (until Section VI), 
$R_f^+ =R_f^- =R_f =  |\overline{A}_{ f } /A_{f }| =  {A}^T_{\overline  f } /A^T_{f }$, as in Eq. (\ref{eq:lambdanonCP}),
so that
\beqa
y^{\prime\pm} &=&  \left(+\left|{q\over p }\right|,-\left|{p\over q} \right|\right)\times (x^\prime _{\overline f}\, \sin\phi \mp y^\prime_{\overline f}\, \cos\phi ),
\nonumber \\ \label{eq:definitions1}
x^{\prime\pm} &=& \left(+\left|{q\over p }\right|,-\left|{p\over q} \right|\right)\times( x^\prime_{\overline f}\, \cos\phi  \pm y^\prime _{\overline f}\, \sin\phi),
\eeqa
where $x^{\prime}_{\overline f}$ and $y^\prime_{\overline f}$ have been defined in Eq.~(\ref{eq:xprimeyprime}).
In addition, 
\beqa |q / p|^{\pm 2} (x^2 + y^2) = (x^{\prime \pm})^2 +  (y^{\prime \pm})^2,
\eeqa
allowing $|q/p|$ to be expressed solely in terms of $(x^{\prime \pm })^2+ (y^{\prime \pm})^2$ above.
The expressions given in Refs.~\cite{Bergmann:2000id} and \cite{BelleKPi} for 
$y^{\prime \pm}$ and $x^{\prime \pm}$ 
differ from those in Eq.~(\ref{eq:definitions1}) due to choice of convention, and are recovered
by substituting $\Delta_f \to -\Delta_f + \pi$ in $x^{\prime}_{\overline f}$ and $y^\prime_{\overline f}$.   
The time dependence for $D^0$ decays to the ``right-sign" (RS) final states is, to good approximation, exponential and given by
\beqa\label{eq:GammahatKPI}
\Gamma[D^0 (t) \to f] &=&e^{-\tau } |A_{f}|^2 ,\nonumber\\
\Gamma[\overline{D^0 }(t) \to \overline{f}]&=&e^{-\tau } |\overline{A}_{\overline{f}}|^2.
\eeqa
Thus, the decay rate parameter is $\hat \Gamma_{D^0 \to K^- \pi^+ } = \Gamma_D$.

A fit to the time-dependence in Eqs. (\ref{eq:DKppim}) and (\ref{eq:DbarKmpip}) yields measurements of $R_f^\pm$, $y^{\prime \pm}$, and $x^{\prime \pm}$,
which can be used to determine or constrain $1-|q/p|$ and $\phi$, as carried out in \cite{BelleKPi} for $D^0 \to K^\pm \pi^\mp$. 
Note that the CP violating quantity $(y^{\prime +} - y^{\prime -} )$ satisfies
\beqa\label{eq:ypymvsDeltaY}
R_{\overline f}\, (y^{\prime +} - y^{\prime -} )=  \Delta Y_{\overline f} \,,
\eeqa 
where $R_{\overline f} = R_f^{-1}$ is the magnitude of the CF to DCS amplitude ratio (for $r_f = r_{\overline f} = 0$),
see Eqs.~(\ref{eq:lambdanonCP}) and (\ref{eq:defdy-non})--(\ref{eq:xprimeyprime}).
Finally, the contributions of CPVMIX and CPVINT in $D^0$ decays to RS final states are 
relatively suppressed by $\tan^4 \theta_c$, and are therefore not considered.


An important CP conserving quantity $y_{\rm CP}$, mentioned in Section II, can be defined in terms of the decay rate parameters $\hat \Gamma_{D^0 \to f_{\rm CP}}$ (for 
SCS decays to CP eigenstates) and  $\hat \Gamma_{D^0 \to K^- \pi^+}$,
\beqa\label{eq:yCPdef} y_{CP}=\eta^{\rm CP}_f \,{\hat\Gamma_{D^0\to f_{CP}} +\hat\Gamma_{\overline{D^0}\to f_{CP}}  \over 2\, \hat\Gamma_{D^0 \to K^- \pi^+ }   } -1 \,.\eeqa
The expressions for the decay rate parameters given above (in the $r_f= 0$ limit) imply \cite{Bergmann:2000id}
\beqa\label{eq:yCPqovpsinphi}
y_{\rm CP}={ y\over 2 } \cos\phi \left(\left|{q\over p}\right|+\left|{p\over q}\right|\right) -{x\over 2 } \sin\phi \left(\left|{q\over p}\right|-\left|{p\over q}\right|\right).
\eeqa


The time-integrated CP asymmetry for $D^0$ decays to CP eigenstates (SCS and CF/DCS) is defined as 
\beqa \label{eq:timeintegrated}
{ a}_f  \equiv {\Gamma (D^0 \to f) - \Gamma (\overline{D^0} \to f) \over \Gamma (D^0 \to f) + \Gamma (\overline{D^0} \to f) }\,.
\eeqa
Expanding to leading order in $x$, $y$, $r_f$
yields~\cite{Grossman:2006jg}
\beqa \label{eq:aftimeintegrated}  a_f = a_f^d + a^m + a^i \,,\eeqa
where $a^m$ and $a^i $ are given in Eq. (\ref{eq:amai}), and
\beqa \label{eq:directCP}
a_f^d = 2 r_f \sin\phi_f \sin\delta_f 
\eeqa
is the (non-universal) direct CP violation contribution.
The time-dependent CP asymmetry ($\Delta Y_f $) and the time-integrated CP asymmetry ($a_f$) are equal if there is no direct CP violation.


For SCS $D^0$ decays to non-CP eigenstates there are two time-integrated
CP asymmetries to consider,
\beqa\label{eq:timeintegratedCP}
a_f\equiv\frac {\Gamma({D}^0\to f)-\Gamma(\overline{D^0}\to\overline{f})}
{\Gamma({D}^0\to f)+\Gamma(\overline{D^0}\to\overline{f})}\,,\no\\
a_{\overline{f}}\equiv\frac
{\Gamma({D}^0\to\overline{f})-\Gamma(\overline{D^0}\to{f})}
{\Gamma({D}^0\to\overline{f})+\Gamma(\overline{D^0}\to{f})}.
\eeqa
Expanding to leading order in $x$, $y$, $r_f$, $r_{\overline f}$ yields~\cite{Grossman:2006jg}
\beqa \label{defamandai}
a_f=a_f^d+a_f^m+a_f^i,~~~~
a_{\overline f}=a_{\overline f}^d+a_{\overline f}^m+a_{\overline f}^i,
\eeqa
where 
$a_{f}^m$, $a_{\overline{f}}^m$ and $a_{f}^i $, $a_{\overline{f}}^i $ are given in Eq. (\ref{eq:amainonCP}), and 
\beqa \label{eq:directCPnonCP}
a_f^d = 2 r_f \sin\phi_f \sin\delta_f ,~~~a_{\overline f}^d = 2 r_{\overline f} \sin\phi_{\overline f} \sin\delta_{\overline f} 
\eeqa
are the direct CP violation contributions.
Again, if there are no new weak phases in decay, the time-dependent and time-integrated CP asymmetries are equal, i.e., $\Delta Y_f = a_f  $ and $\Delta Y_{\overline f} = a_{\overline f}  $.    

In the case of CF/DCS decays to non-CP eigenstates, and in our convention for RS and WS final states,
the definitions of $a_f$ and $a_{\overline f}$ in Eq. (\ref{eq:timeintegratedCP}) correspond to the RS and WS time-integrated CP asymmetries, respectively
(e.g., $a_{K^- \pi^+} $ and $a_{K^+ \pi^-}$ for $D^0 \to K^\mp \pi^\pm$).
To leading order in $x$, $y$, $r_f$, and $r_{\overline f}$ they are given by
\beqa \label{eq:timeintegCFDCS}
a_f = a_f^d,~~~~~~~a_{\overline{f} } = R_{\overline{f} }\,(y^{\prime +} - y^{\prime -} ) + a_{\overline f}^d,
\eeqa
where the RS ($a_f^d$) and WS ($a_{\overline f}^d$) direct CP asymmetries are as in Eq. (\ref{eq:directCPnonCP}).

The time-dependent CP asymmetry 
for $B_s$ decay to a CP eigenstate, to leading order in $r_f$ and for $|q/p|=1 $ (the HFAG average is $|q/p| =1.002 \pm 0.005$ \cite{HFAGcharm}), 
takes the simple form~\cite{nirlectures}
\beqa\label{eq:Bstimedep}
&&\hspace{-1cm}{\Gamma (B_s (t) \to f ) - \Gamma(\overline{B_s} (t) \to f ) \over \Gamma (B_s (t) \to f )+ \Gamma(\overline{B_s} (t) \to f )}\no\\
&&~~~~~~~~~~~~= S_f \sin( |x| \Gamma t) - C_f \cos(|x| \Gamma t)\,,
\eeqa
where
\beqa\label{eq:SfCP}
S_f =\eta_f^{CP}  {\rm sign}(x) \sin \phi \,,~~~C_f =2 r_f \sin \phi_f  \sin \delta_f 
\eeqa
are the contributions due to interference between mixing and decay, and
direct CP violation, respectively.
The factor ${\rm sign}(x)$ in $S_f$ originates from the time-dependence of the decay rates, via
$\sin(x \Gamma t ) = {\rm sign (x)} \sin(|x| \Gamma t)$, and insures that $S_f$ is independent of sign convention.

For $B_s $ decays to non-CP eigenstates there are two time-dependent CP asymmetries to consider,
\beqa\label{eq:Bstimedep}
&&\hspace{-1cm}{\Gamma (B_s (t) \to f ) - \Gamma(\overline{B_s }(t) \to \overline f ) \over \Gamma (B_s (t) \to f ) + \Gamma(\overline{B_s} (t) \to \overline f )}\no\\
&&~~~~~~~~~~~~= S_f \sin( |x| \Gamma t) - C_f \cos(|x| \Gamma t)\,,\no\\
&&\hspace{-1cm}{\Gamma (B_s (t) \to \overline f ) - \Gamma(\overline{B_s} (t) \to f ) \over \Gamma (B_s (t) \to \overline f ) + \Gamma(\overline{B_s} (t) \to f )}\no\\
&&~~~~~~~~~~~~= S_{\overline f} \sin( |x| \Gamma t) - C_{\overline f} \cos(|x| \Gamma t)\,,\eeqa
where (again to leading order in $r_f$, and for $|q/p|=1 $),
\beqa\label{eq:SfnonCP}
S_f = S_{\overline f} = 2  \,{\rm sign}(x)  \sin(\phi)  \cos (\delta_f ) \, r_f /(1+ r_f^2 )\,,\no  \\
C_f = 2 r_f \sin \phi_f \sin \delta_f\,,~~~~C_{\overline f }= 2 r_{\overline f} \sin \phi_{\overline f} \sin \delta_{\overline f}\,. \eeqa
The equality between $S_f$ and $S_{\overline f}$ holds, up to negligible corrections of $O(|q/p|-1)$.

\section{Relating the indirect CP asymmetries}\label{sec:indirectCPasymmetries}
In general, we are interested in decays to final states common to $M^0$ and $\overline{M^0}$,
whose leading contributions to $\Gamma_{12}$ are proportional to the dominant CKM structure entering this quantity, i.e.,
$(V_{cs} V^*_{us})^2 $ for the $D^0$ and $(V_{cb} V_{cs}^* )^2 $ for the $B_s$.  All of the examples we have mentioned previously
are in this class.  In this section we assume that there are no subleading amplitudes with new weak phases in these decays [$r_f = r_{\overline f} =0$ in Eq.~(\ref{eq:fouramp})], and we 
neglect CKM suppressed contributions to $\Gamma_{12}$.  The following relations are then satisfied:
\beqa\label{eq:nodirectCP} 
{ \Gamma_{12}\over \Gamma_{12}^* } = {\overline{A}_f A_f^* \over \overline{A}_f^* A_f } =  \left({ \overline{A}_f \over A_f }\right)^2
\eeqa
and 
\beqa\label{eq:nodirectnonCP} 
{ \Gamma_{12}\over \Gamma_{12}^* } = {\overline{A}_f A_f^* +  \overline{A}_{\overline f} A_{\overline  f}^*     \over
\overline{A}_f^* A_f +  \overline{A}_{\overline f}^*  A_{\overline  f}   } ={ \overline{A}_f \over A_f  }{ \overline{A}_{\overline f} \over A_{\overline f} }\,,  \eeqa
 for CP-eigenstate and non-CP eigenstate final states, respectively.
CKM suppressed contributions to $\Gamma_{12}$ and to $r_f ,\,r_{\overline f}$ within the SM yield corrections to these relations of $O(|(V_{cb} V_{ub}) /(V_{cs} V_{us}) |)\approx 6 \cdot10^{-4}$ for $D^0$ decays, and of $O(|(V_{ub} V_{us}) /(V_{cb} V_{cs}) |)\approx 0.02$ for $B_s$ decays [see Eq. (\ref{eq:deltaphiSM})].
 
The following formulae, obtained from 
Eqs.~(\ref{eq:eigenvalues}) and~(\ref{eq:qoverp}), will be useful:
\beqa \label{eq:sinphi12}
\hspace{-.4cm}
|{q / p}|^2 (x^2+y^2) = x_{12}^2  +y_{12}^2 + 2 x_{12} y_{12} \sin \phi_{12}  \,,\\
\label{eq:qovp4}
\left|{q\over p}\right|^4 =  \left({x_{12}^2 + y_{12}^2 + 2 x_{12} y_{12} \sin \phi_{12} \over  
x_{12}^2 + y_{12}^2 - 2 x_{12} y_{12} \sin \phi_{12} } \right)   \,,\\
\label{eq:y12andx12}
y_{12}^2 = {y^2 + A_m^2 x^2  \over 1- A_m^2 }, ~~~~x_{12}^2 = {x^2 + A_m^2 y^2 \over 1-A_m^2}\,,
\eeqa
where 
\beqa A_m \equiv  (|q/p|^2 -1 )/(|q/p|^2 + 1 )
\eeqa 
is related to CP violation in mixing.  Note that Eq. (\ref{eq:qovp4}), which also appears in \cite{Bigi:2009df}, 
relates CPVMIX to 
the underlying mixing parameters $x_{12}$, $y_{12}$, and $\phi_{12}$.

Multiplying (see Eq. (\ref{eq:qoverp}))
\beqa \left({q\over p }\right)^2 = {M_{12}^* -{i\over 2} \Gamma_{12}^*   \over
M_{12} -{i\over 2} \Gamma_{12} } \label{eq:qovp2}\eeqa
on the l.h.s.~by $(\overline{A}_f /A_f )^2 $ for decays to CP eigenstates to obtain $\lambda_f^2$ 
(or by $(\overline{A}_f \overline{ A}_{\overline f}) /(A_f A_{\overline f}) $ for decays to non-CP eigenstates to obtain $\lambda_f \lambda_{\overline f}$), and on the r.h.s. by $ \Gamma_{12}/ \Gamma_{12}^*$
yields
\beqa
\tan 2\phi &=&  -{\sin 2 \phi_{12} \over \cos 2\phi_{12}  + y_{12}^2 /x_{12}^2 } \label{eq:fiteqphi}\,,\\
\label{eq:sin2phicos2phi} \hspace{-0.6cm}\sin 2 \phi   =&&\hspace{-.6cm}- {2 A_m  x y \over y^2 + A_m^2 x^2 }\,,~~~\cos 2\phi = {y^2 -A_m^2 x^2 \over y^2 + A_m^2 x^2 }\,.
\eeqa
The first relation is incorporated into the fit of $x_{12}$, $y_{12}$, and $\phi_{12}$ using the $D^0 - \overline{D^0}$  mixing
data.
The last two relations are obtained by eliminating the dependence of $\sin 2\phi$ and $\cos2\phi$ on $x_{12}$, $y_{12}$, and $\phi_{12}$, using
Eqs.~(\ref{eq:eigenvalues}),~(\ref{eq:sinphi12}--\ref{eq:y12andx12}).  Finally, a trigonometric identity yields 
\beqa  \tan\phi = -A_m   x /y \,.\label{eq:modindepreln}\eeqa
This expression also appears in \cite{Ciuchini:2007cw}.
It relates CPVMIX to CPVINT, model-independently,
in decay modes in which there are no new weak phases, and is 
independent of sign convention for $x$ or $y$.
In the limit $||q/p| -1|<<1$, which holds to very good approximation for all four meson system, we obtain
\beqa \label{eq:modindeprelnapprox}
\tan\phi = \left(1-\left|{q \over p}\right|\right)  {x\over y}  =- {a_{\rm SL}\over 2 }   {x\over y}  \,.
\eeqa
As discussed in~\cite{Grossman:2009mn}, this relation gives an excellent
description of the data in the neutral kaon system.

It is straightforward to relate $\Delta Y_f$ and the semileptonic CP asymmetry using Eq.~(\ref{eq:modindepreln}), after expanding to first order in $|q/p| -1$.
In the case of $D^0$ decays, the same relations also apply to the time-integrated CP asymmetries (for $r_f = r_{\overline f}=0$).
For decays to CP eigenstates, one obtains
\beqa \Delta Y_f = a_f  =  - y \cos\phi  \,  \eta_f^{CP} \, {a_{\rm SL}\over 2 }  {y^2 + x^2 \over y^2 }\,.
\label{eq:DeltaYfaSLD0} \eeqa
We know from experiment that the level of CP violation in the $D^0 $ system 
is small and that the short-lived meson is approximately CP-even, implying
$|\cos\phi |\approx 1$, ${\rm sign}(y \cos\phi) = +1$ (as in the Standard Model) and, to good approximation,
\beqa \Delta Y_f = a_f = - \eta_f^{CP}  {a_{\rm SL} \over 2 }  {y^2 + x^2 \over |y| }\,,
\label{eq:DeltaYfaSLD0final} \eeqa
which is independent of sign convention for $x$ or $y$.
Similarly, we obtain
\beqa\label{eq:yCPaSL}
y_{\rm CP } = y/\cos\phi = |y| \,,
\eeqa
up to corrections of order $\sin^2 \phi $ or $a_{\rm SL}^2 $.

For SCS $D^0$ decays to non-CP eigenstates,  one obtains
\beqa 
\label{eq:Yfequalities} R_f \Delta Y_{\overline f} &=& \Delta Y_f / R_f = R_f \,a_{\overline f} = a_{f} /R_f \nonumber\\
&=&-\cos \Delta_f \, {a_{\rm SL}  \over 2} {y^2 + x^2  \over |y|}\,.
\eeqa
Confirmation of the relation between the hadronic CP asymmetries in the first line of Eq. (\ref{eq:Yfequalities})
does not require knowledge of $\Delta_f$.  
In terms of the CP-averaged branching ratios for $D^0 \to f$ and $D^0 \to \overline f$ decays, it is simply given by
\beqa \label{eq:DeltaYBRreln}
{ \Delta Y_{\overline  f} \over \Delta Y_f } = {a_{\overline f} \over a_f } = {{\rm Br} (D^0 \to f ) \over {\rm Br} (D^0 \to \overline f) }\,.
\eeqa
This relation follows non-trivially from Eq.~(\ref{eq:modindepreln}): $R_f \Delta Y_{\bar f}$ (or $R_f \,a_{\overline f}^{CP } $) and $ \Delta Y_f /R_f $ (or $a_{f}^{CP} /R_f $)
could, in principle, differ by $O(1)$ given that $y\sim x$, that $\sin\Delta_f $ could be large, and that $(|q/p| -1 )\sim \sin\phi $ is allowed.

In the case of CF/DCS $D^0$ decays to non-CP eigenstates, Eqs.~(\ref{eq:ypymvsDeltaY}),~(\ref{eq:timeintegCFDCS}),~and~(\ref{eq:Yfequalities}) 
imply that the time-dependent, time-integrated, and semileptonic CP asymmetries are related as
(recall $f =K^- \pi^+ $ for $D\to K\pi$ in our convention)
\beqa \label{eq:ypymaSL} y^{\prime +} - y^{\prime -}  = R_{f} a_{\overline f} =  -\cos \Delta_f \, {a_{\rm SL} (D^0) \over 2} {y^2 + x^2  \over |y|}\,.\eeqa
The strong phase $\Delta_{K\pi}$ for $D^0 \to K^\pm \pi^\mp$ decays
can be precisely measured by the BES-III Collaboration at the $\Psi ( 3770)$ charm threshold.

For $B_s $ decays to CP eigenstates, the time-dependent and semileptonic CP asymmetries are related as
\beqa\label{eq:SfaSLinBs}
\hspace{-0.2cm}{ 2 S_f /(1- S_f^2 )^{1/2} } = -\eta_f^{CP}  {\rm sign}(y \cos\phi )  { a_{\rm SL}} {|x / y|}  \,, 
\eeqa
which is independent of sign convention for $x $ or $y$
($|x|$ follows from ${\rm sign} (x)$ in $S_f$). 
At this point, we elaborate on the   
determination of ${\rm sign}(y \cos\phi)$ in~\cite{Grossman:2009mn}.
Starting with Eq.~(\ref{eq:qoverp}), the last relation in Eq.~(\ref{eq:eigenvalues}), and taking $y\ll x$
(using the HFAG averages \cite{HFAGcharm}, the central value for $y/x$ is $\approx 0.003$),
we obtain
\beqa \label{eq:ycosphiphysics}
\hspace{-0.3cm}{\rm sign}(y \cos\phi ) &=& {\rm sign}( |M_{12}|^2  {\rm Re}[ \Gamma_{12}^* \bar A_f /A_f  ]+\nonumber\\
&&{\rm Im}[M_{12} \Gamma_{12}^*]\,{\rm Im}[M_{12}^* \bar A_f/A_f])\,.
\eeqa
The ratio of second to first terms above is given by
$\sin \phi_{12} \sin\phi /\cos(\phi_{12} + \phi )$.  However, for $y\ll x $,
$\phi_{12} = \phi\, {\rm mod}(\pi) $, see Eq. (\ref{eq:fiteqphi}), implying that the magnitude of the ratio is less than 1.
Thus, Eq.~(\ref{eq:ycosphiphysics}) simplifies, and
\beqa \label{eq:ycosphiphysicsfinal}
{\rm sign}(y \cos\phi ) = {\rm sign}(   {\rm Re}[ \Gamma_{12}^* \bar A_f /A_f  ])\,.
\eeqa
Given that the impact of new physics on the r.h.s.
would in general be subleading, we conclude that
${\rm sign} (y \cos\phi ) = {\rm sign} (y \cos\phi )_{\rm SM} = +1$, and that
\beqa\label{eq:SfaSLinBs}
\hspace{-0.2cm}{ 2 S_f /(1- S_f^2 )^{1/2} } = -\eta_f^{CP}  {|x / y|} \,{ a_{\rm SL} } \,
\eeqa
in the absence of new weak phases in decay, as in~\cite{Grossman:2009mn}.

For decays to non-CP eigenstates, 
\beqa\label{eq:SfaSLnonCP}
S_f = S_{\overline f} = - \kappa \, |x/y| \,a_{\rm SL} /2\,,
\eeqa
where 
\beqa\label{eq:kappa}
\kappa = (4\,R_f^2 \cos^2 \Delta_f - S_f^2 \,(R_f^2 + 1)^2 )^{1/2} / (R_f^2 + 1)\,,
\eeqa 
and, as usual, $R_f = |\overline{A}_f / A_f |= A_{\overline f}^T / A_f^T $.
The (near) equality of the two time-dependent CP asymmetries, already noted in 
Eq.~(\ref{eq:SfnonCP}), is a trivial consequence of 
$y\ll x $ and $||q/p| -1| \ll1$, unlike in $D^0$ decays.

\section{Fit results for $D^0 - \overline {D}^0$ mixing}\label{sec:Dfits}
The current $D^0 - \overline {D}^0 $ mixing and CP violation 
fit results reported by
the Heavy Flavor Averaging Group (HFAG)
~\cite{HFAGcharm} can be expressed in terms of 
the four universal parameters
($x$, $y$, $|q/p|$, and $\phi$),
two strong phases ($\delta_{K\pi}$ and $\delta_{K\pi\pi}$), 
the CP averaged ratio of wrong-sign to right-sign $D^0 \to K^\pm \pi^\mp$ 
decay rates ($R_D$), and the 
corresponding direct CP violation parameter ($A_D$).
In terms of our notation for CF/DCS decays, 
\beqa \label{eq:AD}
&R_D&= {(R^+_{f})^2  + (R^-_{f} )^2 \over 2}\,, \nonumber\\
A_D &=& {(R^+_{f})^2  - (R^-_{f} )^2  \over (R^+_{f})^2  + (R^-_{f} )^2 }=a_f^d  - a_{\overline f}^d \,,
\eeqa
with $f = K^- \pi^+$.
The four universal parameters are extracted from fits to 
the time-dependent decay rates for 
$D^0 \to K^+ K^- , \pi^+ \pi^- , K^\pm \pi^\mp, K \pi\pi$, and the
semileptonic decay rates~\cite{HFAGcharm}.
The HFAG fit only allows for new weak phases in the $D^0 \to K^\pm \pi^\mp$ amplitudes, via $A_D\ne 0$.
New weak phases in decay and their impact on the $D^0 - \overline {D}^0 $ mixing and CP violation 
fit are discussed in more detail in Sec.~VI.


 \begin{table}[!tb]
\begin{center}
\begin{tabular}{cccc}
\hline
$x \,[\%]$ & $y\,[\%]$ & $|q/p|$ & $\phi \,[{\rm rad}]$\\
\hline
$1.00 \pm 0.25 $    &  $0.77\pm 0.18 $  &
 $0.94\pm 0.14 $ & $-0.046 \pm 0.093 $ \\ 
  $ 1.00 \pm 0.25$ & $0.76 \pm   0.18$ & $0.86 \pm 0.16$ & $ -0.15 \pm  0.13 $\\  \hline
   $\delta_{K\pi}  \,[{\rm rad}]$ & $\delta_{K\pi\pi}  \,[{\rm rad}] $& $ R_D \,[\%]$  & $A_D  \,[\%]$  \\ \hline
   $  0.40 \pm 0.19 $ & $0.20 \pm 0.37$ &$0.336\pm 0.008 $ & 0 \\
   
  $0.39\pm 0.18$ &$0.20 \pm 0.37$  & $0.336\pm 0.009 $& $ -2.1 \pm 2.4$
    \\  \hline
\end{tabular}
\end{center}
\vspace{-0.2cm}
\caption{HFAG outputs for $A_D = 0$ (first row) and $A_D \ne 0$ (second row)}
\label{tab:HFAGoutputs}
\end{table}

 \begin{table}[!tb]
\begin{center}
\begin{tabular}{cccc}
\hline
Parameter & $A_D = 0 $ & Eq.~(\ref{eq:fiteqphi}) removed &\hspace{-0.470cm}\vline  $\,A_D \ne 0$\\
\hline
$x_{12}~ [\%]$                            &  $1.00 \pm 0.25 $   & $1.00 \pm 0.25 $   &  \vline $\,1.02 \pm 0.24 $    \\ 
$y_{12} ~[\%]$         &   $0.77 \pm 0.18$   &   $0.78 \pm 0.18  $ &\vline  $\,0.75 \pm 0.18$\\ 
$\phi_{12}~[{\rm rad}] $                & $0.02 \pm 0.08$       & $-0.12 \pm 0.22$  &\vline $\,0.07 \pm 0.08$ \\ \hline
\end{tabular}
\end{center}
\vspace{-0.2cm}
\caption{Results for the mixing parameters at $1\sigma$, see Sec.~V  (VI) for $A_D= 0$ ($A_D \ne 0$).}
\label{tab:outputs}
\vspace{-0.2cm}
\end{table}

In general, $x$, $y$, and $|q/p|$ can be expressed in terms of the mixing parameters
 $x_{12}$, $y_{12}$, $\phi_{12}$, see Eqs.~(\ref{eq:xfiteq}),~(\ref{eq:yfiteq}),~(\ref{eq:qovp4}).
In the absence of new weak phases in decay, the same is true for $\phi$, see Eq.~(\ref{eq:fiteqphi}).
Using these four equations (recall that Eqs.~(\ref{eq:xfiteq}) and~(\ref{eq:yfiteq}) correspond to the HFAG convention, which identifies $M_2$ with 
the approximately CP-even state) $x$, $y$, $|q/p|$, and $\phi$ 
are determined by the mixing parameters $x_{12}$, $y_{12}$, $\phi_{12}$.
Ranges for these underlying parameters can be
extracted directly from experimental data under the assumption
that $ A_D = 0 $;
where HFAG currently reports seven parameters, only six would be reported.

For this work, we adopt a simpler strategy for extracting
values of  $x_{12}$, $y_{12}$, and $\phi_{12}$:
we take the HFAG fit results (for the $A_D =0$ case) for
$x$, $y$, $|q/p|$, $\phi$, $\delta_{K\pi}$, $\delta_{K\pi\pi}$, $R_D$, 
shown in Table \ref{tab:HFAGoutputs},
and minimize
\begin{equation}
\chi^2 = \epsilon_i W_{ij} \epsilon_j
\label{eq:chiSqaureEqn}
\end{equation}
where $ \epsilon_i $ is the difference between the
HFAG value for the $ i^{th} $ parameter and the fitted
value predicted using the equations which relate
$x$, $y$, $|q/p|$, and $\phi$  to
$x_{12}$, $y_{12}$, and $\phi_{12}$;
the weight matrix $ W_{ij} $ is the inverse of
the full error matrix for the  values reported
by HFAG, including the correlation coefficients~\cite{AJS-private}.
The fitted values for $\delta_{K\pi}$, $\delta_{K\pi\pi}$, and $R_D$
are very close to the HFAG values;
they change only due to (small) off-diagonal elements in $ W_{ij} $.
The HFAG parameters used as input  are taken from a fit with
$ \chi^2 = 24.9 $ for 21 degrees of freedom (28 experimental results
minus 7 parameters).
The value of $ \chi^2 $ in our fit is 0.2.
In effect, the overall $ \chi^2 $ increases slightly 
as one degree of freedom is restored to the mix of
measurements and the parameters to be extracted.

The fitted values of  $x_{12}$, $y_{12}$, and $\phi_{12}$
are listed in the second column of Table~\ref{tab:outputs}.  
In particular, we obtain, 
\beqa \label{eq:phi12bnd} \phi_{12}^D~[{\rm rad}] =0.02 \pm 0.08~({\rm 1\,\sigma})\,.
\eeqa
Our results for $x_{12}$ and $y_{12}$ are very close to the fitted values for $x$ and $y$ in Table~\ref{tab:HFAGoutputs}, as would be expected for
small $\phi_{12}$, see Eqs.~(\ref{eq:xfiteq}),~(\ref{eq:yfiteq}).
A bound equivalent to 
a precision on $\phi_{12}^D$ of $\pm 0.18$ ($1\sigma$), which assumes no correlations between the experimental measurements, has recently been obtained in \cite{Gedalia:2009kh}.
The HFAG error matrix corresponds to parabolic errors, and thus our
two sigma and higher CL intervals are simple multiples
of our $1\sigma$ CL interval.  However, a preliminary HFAG fit~\cite{alantalk} 
to the data, which uses Eqs.~(\ref{eq:xfiteq}),~(\ref{eq:yfiteq}),~(\ref{eq:qovp4}), and~(\ref{eq:fiteqphi}),
as discussed above, indicates that the errors on $\phi_{12}$ are non-parabolic (and thus we do not list
higher-CL intervals).
Therefore, our fit result for $\phi_{12}$ is only approximate.  The preliminary HFAG $1\sigma$ and 95\% CL intervals for non-parabolic errors 
are
\beqa\label{eq:HFAG1sigma}
\phi^D_{12}  \,[{\rm rad}]&=&0.02^{+0.06}_{-0.13}~~(1\sigma)\,,\nonumber\\
&\in& [-0.30,+0.30]~~({\rm 95\%\,CL})\,.
\eeqa
The former is similar to our result using parabolic errors.
The HFAG fit results for parabolic errors are in agreement with ours.

The impact of the relation between $\phi$ and $\phi_{12}$ on the precision with which 
$\phi_{12}$ is constrained is seen by repeating the fit for the $A_D =0$ case, but with Eq.~(\ref{eq:fiteqphi}) removed.
In this case $\phi$ is treated as an independent parameter which is trivially fit.
The result is reported in the third column of  Table~\ref{tab:outputs}.
We observe that the error on $\phi_{12}$ increases by roughly a factor of three, and thus conclude that
the relationship between CPVMIX and CPVINT provides a powerful constraint
on the allowed magnitude of CP violation in $D^0 - \overline{D^0}$ mixing.

Finally, to understand the implications of the bound on $\phi^D_{12} $ for model building, 
we separate $M_{12}$ into its SM and new physics parts,
\beqa  
M_{12} = M_{12}^{\rm SM} e^{i \phi_M^{\rm SM} }+ M_{12}^{\rm NP} e^{ i \phi_M^{\rm NP}}\,,
\eeqa
where only the difference of the weak phases $\phi_M^{\rm NP} - \phi_M^{\rm SM}$ is physical.
We continue to assume that there are no new weak phases in decay, 
and identify $\Gamma_{12}$ with its SM value.
The definition of $\phi_{12}$ then yields
\beqa \label{eq:sinphi12M}
\sin\phi_{12}^D = \left|{M_{12}^{\rm NP} \over M_{12} }\right| \sin(\phi_M^{\rm NP} -\phi_M^{\rm SM })\,,
\eeqa
where $|M_{12}|$ follows from the fitted value of $x_{12}$.
In the usual phase convention in which $M_{12}^{\rm SM}$ is real ($\phi_M^{\rm SM} = 0$), the above bounds on $\phi_{12}^D$ thus imply that
\beqa \label{eq:ImM12bnd1sigma}  {{\rm Im} (M_{12}^{\rm NP} ) \over |M_{12}  |} \in [-0.06,+ 0.10]~~(1\sigma)
\eeqa
for parabolic errors, and 
\beqa \label{eq:ImM12bnd95CL}   {{\rm Im} (M_{12}^{\rm NP} ) \over |M_{12}  |} &\in& [-0.11,+0.08]~~(1\sigma)\,,\nonumber\\
&\in&[-0.30,+ 0.30]~~(95\%{\,\rm CL})\,.
\eeqa
for the (preliminary) non-parabolic HFAG errors.
As shown in the next section, these bounds can not be substantially altered 
if we allow for new weak phases in decay.

\section{New weak phases in decay}\label{sec:CPcorrections}
\subsection{General considerations}
In this section we discuss how the relations between CPVMIX and CPVINT are modified by new weak phases from subleading 
decay amplitudes (originating from new physics, or CKM suppressed SM amplitudes).
We begin with the resulting shifts in ${\rm arg}(\lambda_f )$, ${\rm arg}(\lambda_{\overline f} )$, and ${\rm arg} (\Gamma_{12}^* /\Gamma_{12})$.         
Expressions relating ${\rm arg}(\lambda_f )$ and ${\rm arg}(\lambda_{\overline f} )$
to $1-|q/p|$, as well as to $\phi_{12}$, which depend on these shifts, are obtained, replacing the previous expressions involving $\phi$.  In turn, new relations between the time-dependent and semileptonic CP asymmetries are 
derived for $D^0 $ and $B_s$ decays.  Direct CP violation bounds are used to 
constrain deviations from the $r_f = r_{\overline f}=0$ case, and the $1\sigma$ intervals for $x_{12}$, $y_{12}$, and $\phi_{12}$ from an appropriately modified fit to the $D^0 - \overline{D^0}$ mixing data are presented.

The argument $\phi_{\lambda_f} \equiv {\rm arg}(-\lambda_f )$ for a decay to a CP eigenstate in Eq. (\ref{eq:lambdaCP})
is shifted,  to first order in $r_f$, as
\beqa\label{eq:lambdaargumentshiftsCP}
\hspace{-0.2cm}\phi_{\lambda_f}= \phi +\delta \phi_{\lambda_f},~~~\delta \phi_{\lambda_f}=  - 2 r_f \cos\delta_f \sin\phi_f .\eeqa
For non-CP eigenstates, the arguments $\phi_{\lambda_f} \equiv {\rm arg}(-\lambda_f  ) $ and
$\phi_{\lambda_{\bar f}}\equiv {\rm arg}(-\lambda_{\overline f} )$ in Eq. (\ref{eq:lambdanonCP}) are shifted by
\beqa\label{eq:lambdanonCPshifts} \delta \phi_{\lambda_f}&=&   - r_f \sin(\delta_f +\phi_f) +r_{\overline f} \sin(\delta_{\overline f} -\phi_{\overline f} )  ,\nonumber \\
\delta\phi_{\lambda_{\overline f}}  &=&     - r_{\overline f} \sin(\delta_{\overline f} +\phi_{\overline f}) +r_{f} \sin(\delta_f -\phi_{f} )  .\eeqa 
The new contribution to ${\rm Arg} (\Gamma_{12} /\Gamma_{12}^* )$ is defined as
\beqa\label{eq:argumentshiftsGamma}
\delta \phi_\Gamma \equiv {\rm arg} \left({\Gamma_{12} \over \Gamma_{12}^* } \right)-
{\rm arg}\left( { \Gamma^0_{12} \over \Gamma_{12}^{0\,* }}\right) \hspace{-0.1cm}= 2\, {\rm Im}\left({\delta \Gamma_{12} \over \Gamma^{0}_{12}}\right)\,,
\eeqa
to leading order in $\delta \Gamma_{12}\equiv\Gamma_{12}  -  \Gamma_{12}^0$, where
$\Gamma_{12}^0$ is the leading SM contribution to $\Gamma_{12}$ proportional to 
$(V_{cs} V^*_{us})^2 $ for the $D^0$ and $(V_{cb} V_{cs}^* )^2 $ for the $B_s$.
Note that $\delta \phi_\Gamma$ is phase redefinition invariant and is an
observable, unlike $ {\rm arg}(\Gamma_{12} )$.

$\delta \phi_\Gamma $ receives contributions from CKM suppressed corrections to $\Gamma_{12}$ within the SM, and from subleading decay amplitudes ($r_f , r_{\overline f} \ne 0$).  (We note that a recent analysis of $\Gamma_{12}$ in the $D^0$ system \cite{Bobrowski:2009zc} indicates that the CKM suppressed corrections to $\delta \phi_{\Gamma}$
could be enhanced from $O(|V_{cb} V_{ub} /V_{cs} V_{us} | )$ in the SM to $O(0.01)$ in models with a fourth family.)
The contribution to $\delta \phi_\Gamma $ from subleading decay amplitudes ($\delta \phi_\Gamma^r $), expressed as a sum over exclusive final states, and to leading order in $r_f ,\,r_{\overline f}$, is given by
\beqa\label{eq:delphiGammaExc} 
 &&-{\delta \phi^r_\Gamma \over 4} \,\big(\sum_{f}  \eta_f^{\rm CP}  (A_f^T)^2 +  \sum_{f,\overline{f}}  2 A_f^T A_{\overline f}^T \cos\Delta_f  \big) =\nonumber \\
&& ~~~~~~~~~~~~\sum_{f}   \eta_f^{\rm CP}  (A_f^T)^2 r_f \cos\delta_f \sin\phi_f  +\hspace{-.2cm} \\
 &&\hspace{-.5cm} \sum_{f, \overline{f}}   A_f^T A_{\overline f}^T \big(r_f \cos(\Delta_f - \delta_f )\sin\phi_f  +  r_{\overline f} \cos(\Delta_{f}  +\delta_{\overline f} )\sin\phi_{\overline f}\big),
\nonumber \eeqa
where the sums are over CP and non-CP eigenstates. 
We learn that $\delta \phi^r_\Gamma $  is of $O(4 {\tilde r}_f \sin\phi_f )$, roughly weighted 
by the fraction of $\Gamma_{12} $ that is attributed to the affected decay amplitudes within the SM.
$\tilde r_f$ is the ``generic" size of $r_f $ and $r_{\overline f}$ in these amplitudes.
The same qualitative conclusion can also be reached via the OPE treatment for $\Gamma_{12}$, in the case of the heavier $B_{d}$ and $B_s$ mesons.


The relation between $\phi$ and $\phi_{12}$ in Eq. (\ref{eq:fiteqphi}) is replaced by
\beqa \!\!\! \tan(2 \phi_{\lambda_f}   -2 \delta \phi_{\lambda_f} + \delta \phi_\Gamma )   =-{{\sin 2 \phi_{12} \over \cos 2\phi_{12}  + y_{12}^2 /x_{12}^2 } }\,\label{eq:fiteqphimod}\eeqa
for decay to a CP eigenstate.
The argument on the l.h.s. is simply $2\phi +  \delta \phi_\Gamma  $, which takes into account the 
shift in ${\rm arg}(\Gamma_{12}/\Gamma_{12}^* )$ in Eq. (\ref{eq:nodirectCP}).
The relation between CPVMIX and CPVINT for decay to a CP eigenstate is now given, in terms of
the observable $\phi_{\lambda_f}$, by 
 \beqa  \tan(\phi_{\lambda_f}   -\delta \phi_{\lambda_f} + \delta \phi_\Gamma /2)   = -A_m   x /y \,.\label{eq:modindeprelnnew}\eeqa
Expanding to lowest order in $r_f$ and $|q/p|-1$ yields
\beqa \label{eq:modindeprelnnewapprox}
\hspace{-.3cm}\tan\phi_{\lambda_f} =\left (1-\left|{q\over p}\right|\right){ x\over y} + {\delta \phi_{\lambda_f} - {\delta \phi_\Gamma / 2} \over \cos^2 {\phi}_{\lambda_f} }\,.\eeqa
Corrections to the relations between the semileptonic and time-dependent CP asymmetries
follow straightforwardly, see below. 

In the case of non-CP eigenstate final states, 
new relations which combine the effects of new weak phases in decays to CP conjugate pairs 
(thus removing the dependence on the strong phase $\Delta _f$) are obtained
by substituting $\phi_{\lambda_f}  \to (\phi_{\lambda_f}+ \phi_{\lambda_{\overline f}} )/2$ and $\delta \phi_{\lambda_f}  \to (\delta\phi_{\lambda_f}+\delta\phi_{\lambda_{\overline f}} )/2 $ in Eqs. (\ref{eq:fiteqphimod})--(\ref{eq:modindeprelnnewapprox}).
In practice, it may be more useful to consider their effect on each decay separately (see the discussion of $D^0\to K^\pm \pi^\mp$ decays below), yielding
\beqa\label{eq:fiteqphimodnonCP} \!\!\! \tan(2 \phi_{\lambda_f} \!\!  &\!\!-\!\!&\!\! 2 \delta \phi_{\lambda_f} - 2 \Delta_f + \delta \phi_\Gamma )   ={{-\sin 2 \phi_{12} \over \cos 2\phi_{12}  + y_{12}^2 /x_{12}^2 } }\nonumber\\
&=&\tan(2 \phi_{\lambda_{\overline f} }  - 2 \delta \phi_{\lambda_{\overline f}} + 2 \Delta_f + \delta \phi_\Gamma )\,
\eeqa
for the dependence of the observables $\phi_{\lambda_f}$ and $\phi_{\lambda_{\overline f}}$ on $\phi_{12}$, and
 \beqa \label{eq:modindeprelnnewnonCP} \tan(\phi_{\lambda_f} \!\! & -&\!\!\delta \phi_{\lambda_f} -\Delta_f + \delta \phi_\Gamma /2)   = -A_m   x /y  \,\nonumber \\
 &=&  \tan(\phi_{\lambda_{\overline f}}   -\delta \phi_{\lambda_{\overline f}} +\Delta_f + \delta \phi_\Gamma /2) \,
 \eeqa
for the modified relations between CPVMIX and CPVINT.


Approximate bounds on $\delta \phi_{\lambda_f}$, $\delta \phi_{\lambda_{\overline f}}$, and $\delta \phi_\Gamma $ 
for $D^0$ and $B_s$ decays 
can be obtained from direct CP violation measurements.
It is instructive to compare them to the current experimental sensitivity 
to $\delta \phi_{\lambda_f}$, $\delta \phi_{\lambda_{\overline f}}$, and $\delta \phi_\Gamma $ 
in time-dependent (mixing-related) measurements.

\subsection{$D^0 - \overline{D^0}$ mixing}
We need to consider new weak phases in singly Cabibbo suppressed (SCS) decays, and their combined effects in Cabibbo favored (CF) and doubly Cabibbo suppressed (DCS) decays.
We begin with a discussion of the former.
The HFAG average for $\Delta Y_f$  \cite{HFAGcharm}, obtained from the BaBar and Belle $D^0 \to K^+ K^-$ and 
$D^0 \to \pi^+ \pi^-$ measurements \cite{BaBarBelleKKPiPi},
is
\beqa \label{eq:DeltaYf}
\Delta Y_f = (-0.123 \pm 0.248)\%\,.
\eeqa
The time integrated CP asymmetries for $D^0 \to K^+ K^-$ and $D^0 \to \pi^+ \pi^-$ are \cite{HFAGcharm},
\beqa
a_{K^+K^-} &=& (-0.16 \pm 0.23 )\%\,,\nonumber\\
a_{\pi^+\pi^-}& =& (0.22 \pm 0.37)\%\,.
\eeqa
The direct CP asymmetries are obtained by subtracting $\Delta Y_f $ from the time integrated CP asymmetries, see Eq.~(\ref{eq:aftimeintegrated}), 
yielding
\beqa\label{eq:adirectexp}
a_{K^+ K^-}^d &=& (-0.04 \pm 0.34)\%\,,\nonumber\\
a_{\pi^+\pi^-}^d &=& (0.34\pm 0.45)\%\,.
\eeqa
(Predictions for $a_f^d$ in the Standard Model suffer from large hadronic uncertainties spanning an order of magnitude or more, and could be as large as $\approx 0.1\%$).
Unless the new physics has a very special structure, 
e.g., parity conserving \cite{kagantalk},  the results for $a_{K^+ K^-}^d$ and $a_{\pi^+ \pi^-}^d$ give rough bounds on the direct CP asymmetries for all decays mediated by 
$c\to u (s \bar s, d\bar d) $ transitions.
Models which can easily produce direct CP asymmetries of this size or larger in SCS decays have been discussed in \cite{Grossman:2006jg}.

In general, strong phase differences enter as $\sin\delta $ in the direct CP asymmetries,
and as $\cos\delta$ in $\delta \phi_{\lambda_{f}}$, $\delta \phi_{\lambda_{\overline f}}$, and $\delta \phi_\Gamma $.
However, the relevant new physics ($\Delta C=1$) effective operators for SCS decays differ from the tree-level SM operators in their color and chirality structures (the QCD penguin operators, most notably the chromomagnetic dipole operator, are relatively
unconstrained by $D^0 -\overline{D^0}$ mixing).
Thus, strong phase suppression is not expected \cite{Grossman:2006jg},
implying that $\delta \phi_{\lambda_f} \sim a_f^d$.
This justifies taking
\beqa \label{eq:phibndsSCS} |\delta \phi_{\lambda_f}|, |\delta \phi_{\lambda_{\overline f}}|, |\delta \phi_\Gamma | \lsim 1\% 
\eeqa
for SCS decays, 
and similarly for the last term in Eq.~(\ref{eq:modindeprelnnewapprox}).

The SCS decays enter the HFAG $D^0-\overline{D^0}$ mixing fit via $\Delta Y_f $ (averaged over $\pi^+ \pi^-$ and $K^+ K^- $) and $y_{\rm CP}$.
In the case of decays to CP eigenstates a new weak phase would shift $\Delta Y_f$, to lowest order in $(1-|q/p|)$ and $r_f$, by
\beqa \label{eq:shiftinYf}
\delta (\Delta Y_f )= - \eta_f^{\rm CP} \,( |y| \, a^d_f  - |x| \,\delta \phi_{\lambda_f} )\,.
\eeqa
This result follows by substituting $\phi \to \phi_{\lambda_f}$ and $|q/p| \to |q \overline{A}_f / p A_f |$ in Eq. (\ref{eq:amai}),
and expanding in small quantities.  Note that the impact of $r_f \sin\phi_f $ is suppressed by mixing ($x,y \sim 10^{-2}$), unlike in $a_f^d$ which enters the time integrated CP asymmetry.  With $a^d_{f },\,\delta \phi_{\lambda_f} < 1\%$ and $x,y \sim 1\%$, we find 
\beqa \label{eq:DeltaYshiftnumber}
|\delta (\Delta Y_f )| \lsim 10^{-4}\,, 
\eeqa
which is less than a few percent of the experimental uncertainty, see
Eq.~(\ref{eq:DeltaYf}).
The shift in $y_{\rm CP}$ due to new weak phases in decay must be even smaller relative to its experimental uncertainty,
given in Eq. (\ref{eq:yCPexp}),
because its dependence on CP violating quantities must be quadratic (and still suppressed by $x$ or $y$). 

The relation between $a_{\rm SL}$ and $\Delta Y_f$ in Eq. (\ref{eq:DeltaYfaSLD0final}) for decays to CP eigenstates is modified, to lowest order in 
$r_f$ and $(|q/p|-1)$, as
\beqa\label{eq:correctedcorreln}
\eta_f^{\rm CP} \Delta Y_f &=&  - {a_{\rm SL}  \over 2 }  {y^2 + x^2 \over |y| } -|y|\, a_f^d  \nonumber\\
& +&|x| (\delta \phi_{\lambda_f } - {\delta \phi_\Gamma/ 2})\,.\eeqa
Given that the approximately CP-even 
$D^0$ mass eigenstate is the shorter-lived and heavier one, 
we have substituted $y\cos\phi \to |y|$ (as before) and $x\cos\phi \to |x|$, and similarly below.
Applied to SCS decays, the new physics correction is again $\lsim 10^{-4} $.
The modified relations satisfed by $a_{\rm SL}$, $\Delta Y_f$, and $\Delta Y_{\overline f}$  for SCS decays to non-CP eigenstates which replace  Eq. (\ref{eq:Yfequalities}) are
\beqa \label{eq:DeltaYfnonCPrelnmods1}
\!\!\!{\Delta Y_f / R_f  + R_f \Delta Y_{\overline f} \over \cos\Delta_f }&=&  - {a_{\rm SL}  }  {y^2 + x^2 \over |y| }- |y|\, (a_f^d +a_{\overline f}^d) \nonumber\\
 &+& |x| (\delta \phi_{\lambda_f} +\delta \phi_{\lambda_{\overline f}} -\delta \phi_{\Gamma} )\,, \\
\!\!\!  \label{eq:DeltaYfnonCPrelnmods2}{\Delta Y_f / R_f  - R_f \Delta Y_{\overline f} \over \sin\Delta_f }&=& |y|\, (\delta\phi_{\lambda_f} + \delta \phi_{\lambda_{\overline f}} -\delta \phi_{\Gamma})\nonumber\\
& +&| x|\, (a_f^d + a_{\overline f}^d )\,,
\eeqa
to lowest order in $r_f$, $(|q/p|-1)$, and neglecting terms of $O(r_f  x \sin\phi  )$, $O(r_f  y \sin\phi )$, and are thus similarly bounded.


The most precisely measured CF and DCS
time-integrated and direct CP asymmetries are near zero, with uncertainties of $\approx 1\%$ and $\approx 5\%$, 
respectively.  For example, the time integrated CP asymmetries for $D^0 \to K^- \pi^+ \pi^0$ (CF) and $D^0 \to K^+ \pi^- \pi^0$ (DCS) are ~\cite{HFAGcharm} 
\beqa\label{eq:CFDCSCPexp}
a_f &=& (+0.16 \pm 0.89)\,\%\,;~~f=K^- \pi^+ \pi^0\,,\nonumber\\
a_f &=&( -1.4 \pm 5.2)\,\%\,;~~f=K^+ \pi^- \pi^0\,.
\eeqa
The difference between the CF and DCS direct CP asymmetries in $D^0 \to K^\pm\pi^\mp$ ($A_D$), averaged over the BaBar and Belle measurements~\cite{BaBarKPi,BelleKPi}, is 
\beqa
a_{K^- \pi^+ }^d - a_{K^+ \pi^-}^d = (0.4\pm 3.5 )\%\,
\eeqa
and the global HFAG $D^0 - \overline{D^0}$ mixing fit gives $(-2.1\pm 2.4   )\%$ (second row, Table I).
Finally, the best CF $D^\pm$ direct CP asymmetry bounds are for $D^+ \to K^- \pi^+ \pi^+ $ and $D^+ \to K_s  \pi^+ \pi^0 $ \cite{HFAGcharm}, 
\beqa\label{eq:chargedDCPexp}
 a^d_f &=& (-0.5 \pm 1.0)\,\%\,;~~ f=K^- \pi^+ \pi^+\,,\nonumber\\
a^d_f &=&( +0.3 \pm 0.9)\,\%\,;~~ f=K_s  \pi^+ \pi^0 \,.
\eeqa
It is difficult to construct models
with non-negligible new weak phases in CF and DCS decays \cite{bergmannir} (one example is known \cite{ambrosio}). 
Again, $\Delta C=1$ effective operators with different color and chirality structures than their SM counterparts would be important, and 
we can expect significant strong phase differences.  The direct and time-integrated CP violation bounds therefore imply
\beqa \label{eq:phibndsCFDCS}
|\delta \phi_{\lambda_f}|, |\delta \phi_{\lambda_{\overline f}}|, |\delta \phi_\Gamma| \le O({\rm few~ percent})
\eeqa
(following our convention, 
take $r_f $ and $r_{\overline f} $ in Eqs.~(\ref{eq:directCPnonCP})~and~(\ref{eq:lambdanonCPshifts}) to correspond to CF and DCS new physics amplitudes, respectively).
For completeness, we note that for CF and DCS decays to CP eigenstates (e.g., $D^0 \to K_s \pi^0,\, \rho^0 K_s$),
$|\delta \phi_{\lambda_f}| \le O(1\%)$, which is the approximate bound on CF direct CP violation (DCS contributions are suppressed by $\theta_c^2$).


New weak phases in CF and DCS transitions would enter the HFAG $D^0-\overline D^0$ mixing fit via 
$D^0 \to K^\mp \pi^\pm$ and $D^0 \to K_s \pi^+ \pi^-$.  
For illustrative purposes, lets consider $D \to K \pi$ in more detail.
The general form for the time-dependent amplitudes $D^0 (t) \to K^+ \pi^- $ and $\overline{D^0}(t)\to K^- \pi^+$
is the same as in Eqs.~(\ref{eq:DKppim})~and~(\ref{eq:DbarKmpip}).  
However, the corrected expressions for $y^\pm$ and $x^\pm$, see Eq. (\ref{eq:definitions1}), are given by ($f=K^- \pi^+ $)
\beqa\label{eq:definitions2}
y^{\prime\pm}\!\! &\!\!= \!\!&\!\!  \left(+\left|{ \overline{A}_{\overline f} \over  A_f}{  q\over p }\right|,-\left|{ {A}_{f} \over  \overline{A}_{\overline f}}{p\over q} \right|\right)\times (x^\prime _{\overline f}\, \sin\phi^\pm \mp y^\prime_{\overline f}\, \cos\phi^\pm ),
\nonumber\\
x^{\prime\pm}\!\! &=&\!\! \left(+\left|{ \overline{A}_{\overline f} \over  A_f}{  q\over p }\right|,-\left|{ {A}_{f} \over  \overline{A}_{\overline f}}{p\over q} \right|\right)\times( x^\prime_{\overline f}\, \cos\phi^\pm  \pm y^\prime _{\overline f}\, \sin\phi^\pm),\nonumber\\
\eeqa
where (in terms of the direct CP asymmetry for the CF decays),  
\beqa \label{eq:CFDCPV} \left|{ \overline{A}_{\overline f}/  A_f}\right|  =  1+ a_f^d = 1+ 2 r_f \sin\delta_f \sin\phi_f \,,
\eeqa
and 
\beqa \label{eq:phipphim}
\phi^+  &=& \phi_ {\lambda_{\overline  f}} +\Delta_f =  \phi + \delta \phi_{\lambda_{\overline f}}\,\nonumber\\
\phi^- &=& \phi_{\lambda_f} -\Delta_f =\phi + \delta \phi_{\lambda_{ f}}\,.
\eeqa
Corrections to the relation between the time-dependent CP asymmetry $(y^{\prime +} -y^{\prime -} ) = R_f \Delta Y_{\overline f}$ and $a_{\rm SL}$ in Eq. (\ref{eq:ypymaSL})
are easily obtained from Eqs.~(\ref{eq:DeltaYfnonCPrelnmods1}) and (\ref{eq:DeltaYfnonCPrelnmods2}), applied to CF/DCS decays.

Measurements of $y^{\prime \pm} $, $x^{\prime \pm}$ and $R_D$, $A_D$ [defined in Eq. (\ref{eq:AD})]
for $D^0 \to K\pi$ have been reported by BaBar and Belle \cite{BaBarKPi,BelleKPi}, also see \cite{HFAGcharm}.
Averaging over the two experiments yields
\beqa \label{eq:yPmyMexp}
y^{\prime +} - y^{\prime -} = (-0.19 \pm 0.64)\%
\eeqa
for the time-dependent CP violation.
The experimental error is an order of magnitude larger than the maximal allowed shift due to new weak phases in decay,  
of order $x$ or $y$ times the bound in Eq. (\ref{eq:phibndsCFDCS}).
In addition, a fit for $x_{\overline f}^\prime $, $y_{\overline f}^\prime$ and $\phi$ has been carried out in the Belle analysis~\cite{BelleKPi},
yielding
\beqa \label{eq:phibndKPi}\phi= (0.16 \pm 0.41)~ [{\rm rad}].
\eeqa  
However, the fit uses the formulae for $x^{\prime \pm }$ and $y^{\prime \pm}$ in Eq. (\ref{eq:definitions1}), 
thus neglecting the corrections in Eq. (\ref{eq:definitions2}).  In particular, it assumes that $\phi^+ = \phi^- = \phi$.  Fortunately, the reported error on $\phi$ is an order of magnitude larger than the upper bounds on $|\delta \phi_{\lambda_{f}} |$ and $|\delta \phi_{\lambda_{\overline f}} |$ of a few percent, in $\phi^\pm$.  Moreover, $a_f^d$ should be $\lsim 1\%$, hence negligible in Eq.~({\ref{eq:CFDCPV}).  Thus, the use of Eq.~(\ref{eq:definitions1}) turns out to be a good approximation.

The Belle Collaboration also fits for $\phi$ in a time-dependent Dalitz plot analysis for $D^0 \to K_s \pi^+ \pi^-$~\cite{BelleKsPiPi},
obtaining 
\beqa\label{eq:phiKsPiPi} 
\phi = (-0.24 \pm 0.32)~ [{\rm rad}]\,.
\eeqa
Again, this analysis assumes that $\phi^+ = \phi^- =\phi$ (in general, $\phi^+$ and $\phi^-$ would vary across the Dalitz plot).
Again, the error on $\phi$ is about an order of magnitude larger than the allowed shifts in $\phi^\pm$, implying that this is a good approximation.

The outputs of the HFAG fit for $A_D \ne 0 $  (new weak phases in decay) listed in Table~I
have been obtained under the assumption that $\phi^+ = \phi^- =\phi$ in 
the time-dependent $D^0 \to K^\pm  \pi^\mp $ and $D^0 \to K_s \pi^+ \pi^- $ amplitudes.
We have just seen that this is a good approximation.
In addition, HFAG has not allowed for new weak phases in $D^0 \to K^+ K^-$ and $D^0 \to \pi^+\pi^- $.
Again, this is a good approximation for SCS decays, given that the
impact of new weak phases on $\Delta Y_f $ and $y_{CP}$ would be negligible.
Finally, modifications to the relation between $\phi$ and $\phi_{12}$ in Eq. (\ref{eq:fiteqphi}) [see  Eqs.~(\ref{eq:fiteqphimod})~and~(\ref{eq:fiteqphimodnonCP})]
are smaller than the experimental sensitivity to $\phi$ in CF/DCS decays by an order of magnitude,
and in SCS decays by more than an order of magnitude.

In view of the above considerations and in the case of new weak phases in decay, the mixing parameters $x_{12}$, $y_{12}$, and $\phi_{12}$ can be obtained, to good approximation, 
along the lines of the fit carried out in Sec.~V (for $A_D=0$).  In particular, $\phi$ is once again expressed in terms of  $x_{12}$, $y_{12}$, and $\phi_{12}$ using Eq.~(\ref{eq:fiteqphi}) [as are
$x$, $y$, and $|q/p|$, using Eqs.~(\ref{eq:xfiteq}),~(\ref{eq:yfiteq}), and~(\ref{eq:qovp4})].  However, 
now we take the HFAG fit results for $A_D \ne 0$, see Table I, and add $A_D$ to the sum over HFAG outputs
in Eq. (\ref{eq:chiSqaureEqn}) for the $\chi^2$ function.
The validity of this approximation reflects the suppression due to mixing ($x$ or $y$) of the effects of new weak phases in decay on CPVINT [continued use of Eq.~(\ref{eq:fiteqphi})],
and the lack of such suppression in the direct CP asymmetries [use of the $A_D \ne 0$ fit results].
The HFAG parameters used as input in this case are taken from a fit with $\chi^2 = 25.3$ for 20 degrees of freedom
(28 experimental results minus 8 parameters). The value of $\chi^2$ in our fit is 1.3. Thus, as in the $A_D=0$ fit, the overall $\chi^2$ increases 
by a small amount as the number of degrees of freedom is increased by one.

The fitted values of $x_{12}$, $y_{12}$, and $\phi^D_{12}$, with $1\sigma$ parabolic errors, are shown in the last column of Table~II.
In particular, we obtain
\beqa \label{eq:phi12fitmodified}
\phi_{12}^D~[{\rm rad}]=  0.07 \pm 0.08~~(1\sigma)
\eeqa
for parabolic errors.
This is fully consistent (within $1\sigma$) with Eq. (\ref{eq:phi12bnd}) for no new weak phases in decay, as expected.  
To ascertain the impact on models in which new weak phases in decay are possible, 
we note that the relation between $\phi_{12}^D$ and $M_{12}^{\rm NP}$ in Eq. (\ref{eq:sinphi12M})
is modified, to very good approximation, as
\beqa \label{eq:sinphi12modified}
\sin\phi_{12}^D = \left|{M_{12}^{\rm NP} \over M_{12} }\right| \sin(\phi_M^{\rm NP} -\phi_M^{\rm SM }) - {\delta \phi_{\Gamma} \over 2}\,.
\eeqa
Therefore, we obtain the approximate (parabolic) $1\sigma$ CL interval
\beqa \label{eq:ImM12bnd1sigmamodified}  {{\rm Im} (M_{12}^{\rm NP} ) \over |M_{12}  |} \in [-0.01,+ 0.15]\,,
\eeqa
(for the usual phase convention in which $M_{12}^{\rm SM} =0$ is real), up to small corrections of 
a few percent or less from $\delta \phi_\Gamma /2$. 
This is consistent, within $1\sigma$, with Eq. (\ref{eq:ImM12bnd1sigma}) for no new weak phases allowed.
Similarly, the corresponding HFAG (non-parabolic error) analysis, i.e., 
a direct fit to the experimental data which allows $A_D \ne 0 $ and incorporates Eq.~(\ref{eq:fiteqphi}), should be consistent 
with Eq. (\ref{eq:ImM12bnd95CL}).

What will the sensitivity to new weak phases in decay be at a high luminosity flavor factory?
We have seen that their impact on the time-dependent CP asymmetries ($\Delta Y_f $) in SCS decays can be at 
most a few percent of the current errors (for $D^0 \to \pi^+ \pi^-$ and $D^0 \to K^+ K^- $).  
In the case of
CF and DCS decays (e.g., $D^0 \to K_s \pi^+\pi^- ,K^\pm \pi^\mp$) 
we saw that their current sensitivity to $\phi$ is roughly an order of magnitude weaker than the maximal shifts allowed in $\phi^\pm$, 
and similarly for the precision with which Eq.~(\ref{eq:ypymaSL}), relating $(y^{\prime +} - y^{\prime -})$ and $a_{\rm SL}$, can be tested.
Thus, even with an order of magnitude reduction in the errors on 
CPVINT and CPVMIX, as might be expected at a super-$B$ factory with $75~{\rm ab}^{-1}$, 
it could be difficult to detect new weak phases in decay at currently allowed levels via time-dependent CP asymmetry measurements.

The effects of new weak phases in decay are much easier to observe in $D^0$ and $D^\pm$ direct CP asymmetry
measurements (in $D^0 $ decays this requires comparison of
the time-integrated and time-dependent CP asymmetries), as they are not suppressed by mixing ($x$ or $y$).  
In particular, there is a good chance of detecting direct CP violation in SCS decays at a super-B factory, 
even if due solely to SM penguins. 
As a further illustration, we observe that 
the sum and difference of CP-conjugate time-integrated CP asymmetries in SCS decays would satisfy
\beqa\label{eq:timeintegrateddiffs}
\!\!\!\!\!\!\!\!{a_f \over  R_f } + R_f a_{\overline f} &=&  {a^d_f \over R_f  }+ R_f a^d_{\overline f}  - {a_{\rm SL}  }  {y^2 + x^2 \over |y| }\cos\Delta_f ,\nonumber\\
\!\!\!\!\!\!\!\!{a_f  \over R_f } &- &R_f a_{\overline f}= {a^d_f \over  R_f } - R_f a^d_{\overline f}\,,
\eeqa
up to negligible corrections of $O(x\, r_{f,\overline f} ) $ and $O(  y \,r_{f,\overline f} )$.
Violations of the $r_{f}=r_{\overline f}=0$ relations satisfied by $a_f $ and $a_{\overline f}$ in Eqs.~(\ref{eq:Yfequalities})~and~(\ref{eq:DeltaYBRreln}) 
could, therefore, be observed at a high luminosity flavor factory  (e.g., in $D^0\to K^* K$ decays) at currently allowed levels, unlike the
violations of the corresponding ($\Delta Y_f  $, $ \Delta Y_{\overline f}$) time-dependent CP asymmetry relations in
Eqs.~(\ref{eq:DeltaYfnonCPrelnmods1}) and (\ref{eq:DeltaYfnonCPrelnmods2}).

Finally, it is interesting to observe that with sufficient statistics it could be possible to isolate and measure new contributions to ${\rm arg}(\Gamma_{12}/\Gamma_{12}^* )$, precisely because the experimental sensitivity to $\phi$ in individual decays substantially lags the direct CP asymmetry errors.
Presumably, when the effects of non-zero values for $(\delta \phi_{\lambda_{f,\overline f} } - {\delta \phi_\Gamma/ 2})$  in Eqs.~(\ref{eq:fiteqphimod})--(\ref{eq:modindeprelnnewnonCP}) are observed,
the direct CP asymmetry errors will be much smaller.  
If, for example, it turns out that 
\beqa\label{eq:gettingphiGamma}
|\delta \phi_{\lambda_f } - {\delta \phi_\Gamma/ 2} | \gg |a_f^d  |\,,
\eeqa
in the case of decay to a CP eigenstate, then we will have obtained a measurement of $\delta \phi_\Gamma$  (since  $\delta \phi_{\lambda_f } \sim a_f^d $).
For example, this situation could be realized: (i) in SCS decays, if new phases in CF/DCS amplitudes are near the current direct CP violation bounds,
or (ii) in CF/DCS decays, in the more likely possibility that new phases only appear in SCS amplitudes.
The determination of $\delta \phi_\Gamma$ could be combined with a measurement of $\phi_{12}^D$ to fix $|M_{12}^{\rm NP}  | \sin (\phi_M^{\rm NP} - \phi_M^{\rm SM})$ in Eq. (\ref{eq:sinphi12modified}).

\subsection{$B_s - \overline{B_s}$ mixing}
Moving to $B_s - \overline B_s$ mixing, the CDF and D0 collaborations are probing $S_f$, in $B_s \to J/\Psi \phi$, and $a_{\rm SL}$ with combined uncertainties of $0.4$ and $0.009$, respectively \cite{HFAGcharm}.  At LHCb with 2$\,{\rm fb}^{-1}$ the expected uncertainties are $\delta S_f \approx 0.02$ \cite{Mainardaspentalk}
and  $\delta a_{\rm SL} \approx 0.002$ \cite{muheimLHCb}.
If new subleading weak phases appear in $B_s$ decays to CP eigenstates, then 
\beqa S_f = \eta_f^{\rm CP}  {\rm sign} (x) \sin \phi_{\lambda_f} \,,\eeqa
with 
$\phi_{\lambda_f} $ given in Eq. (\ref{eq:lambdaargumentshiftsCP}).
The modified relation between $a_{\rm SL}$ and $S_f$ for decays to CP eigenstates [see Eq. (\ref{eq:SfaSLinBs})] follows from Eq. (\ref{eq:modindeprelnnewapprox}), and is given to
lowest order in $r_f$ and $|q/p|-1$ by
\beqa\label{eq:SfaSLinBsmod}
\hspace{-0.2cm} 
\!\!\!\!\eta_f^{CP} \, {  S_f \over (1- S_f^2 )^{1/2} }\hspace{-.5cm} &&= -{\rm sign }(y \cos\phi_{\lambda_f}) {\left| x \over y\right|} \,{ a_{\rm SL}\over 2} \nonumber\\\!\!\!\!&-&\!\!{{\rm sign} (x\cos\phi_{\lambda_f}) \over \cos^2 \phi_{\lambda_f} }\left( {\delta \phi_\Gamma\over  2}-\delta \phi_{\lambda_f } \right)\!.\eeqa
The new physics satisfies the inequality $ |(2 r_f \cos\delta_f \sin \phi_f ) \tan\phi | < 1$ (unless $\phi \approx \pi/2 $), which implies that
${\rm sign }(y \cos\phi_{\lambda_f}) = {\rm sign }(y \cos\phi )=+1$ in the first term, and  ${\rm sign }(x \cos\phi_{\lambda_f}) = {\rm sign }(x \cos\phi )$ in the second term.

There are no direct CP asymmetry measurements yet for $B_s$ decays mediated by $b \to c\bar c s$
transitions.  However, their magnitudes should be of same order as those for 
$b \to c\bar c s$ transitions in $B_d$ decays.
The best bound is $\approx 2\%$, from the direct CP asymmetry for $B_d \to J/\Psi K^0$ \cite{HFAGcharm}.
As previously noted, the strong phase differences enter as $\sin\delta_f $ in the direct CP asymmetries, and as $\cos\delta_f$ in $\delta \phi_{\lambda_f}$
and $\delta \phi_\Gamma $.
However, in the Standard Model the $B\to J/\Psi \phi$ amplitude is given by a particular color-suppressed combination of two effective operator 
matrix elements ($Q_{1,2}$).  Moreover, the soft gluon contributions to these matrix elements are formally suppressed by $\Lambda_{\rm QCD} /(m_b \alpha_s) $ rather than $\Lambda_{\rm QCD} /m_b $
\cite{BBNS1}.  We also note that significant strong phase differences ($\sim 30^\circ - 50^\circ$) between the different isospin amplitudes
in $B\to D^{(*)} \pi$ and $B\to D^{(*)} K$ decays are known to exist, due to color-suppressed channels \cite{Mantry:2003uz}.
Thus, significant
strong phase differences between the SM and new physics $B_s \to J/\Psi \phi$ or $B_d \to J/\Psi K^0$ amplitudes can be expected.
We therefore take 
\beqa \label{eq:phibndsBs}
|\delta \phi_{\lambda_{f}}|,\,|\delta \phi_\Gamma  |\lsim 5\%  \,,\eeqa
for new physics in $b \to c\bar c s$ transitions.
Effects of this size in the second term on the r.h.s. of Eq. (\ref{eq:SfaSLinBsmod}), applied to $B_s \to J/\Psi \phi$, would be difficult to observe at LHCb, given an order of magnitude larger projected experimental uncertainty on the first term
\beqa \delta \left({a_{\rm SL}\over 2} \, \left|{x \over y}\right|\right) = O(0.4)\,,
\eeqa
for 2 ${\rm fb}^{-1}$ (obtained from $\delta a_{\rm SL}$ above, and the SM central value for $|x/y|$ in \cite{lenznierste}).

New CP violating effects at the 5\% level in the tree-amplitudes would be quite exotic.
If the new physics enters the $b \to c\bar c s$ transitions via gluonic or electroweak penguins, which we believe is a far more likely scenario,
then its contributions to $\delta \phi_{\lambda_{J/\Psi \phi } }$ and $\delta \phi_\Gamma $ would be negligible.
Recall that new CP violating amplitudes in penguin dominated $B_d$ decays, e.g., $B_d \to \phi K_s$, are constrained to lie below $O(10\%)$.

Finally, $\delta \phi_\Gamma$ receives
a significant Standard Model contribution (relative to the leading $[\lambda_c /\lambda_c^{* } ]^2 $ CKM structure in $\Gamma_{12}/\Gamma_{12}^* $)
\beqa \label{eq:deltaphiSM}
\delta \phi^{\rm SM}_\Gamma = 4 \,{\rm Im} \!\left({\lambda_u  \over \lambda_c }  \right) \left({\Gamma_{12}^{uc} \over \Gamma_{12}^{cc} }\right)_{\rm SM}\!\!
\approx 8\%  \left({\Gamma_{12}^{uc} \over \Gamma_{12}^{cc} }\right)_{\rm SM} \,,
\eeqa
where 
$\lambda_i \equiv V_{is}^* V_{ib}$.
$\Gamma_{12}^{cc}$ and $\Gamma_{12}^{uc}$ are defined in the Standard Model expression for $\Gamma_{12}$,
\beqa \label{eq:GammaijOPE}
\Gamma_{12}^{\rm SM} = -(\lambda_c^2\, \Gamma_{12}^{cc} + 2 \lambda_c \lambda_u \, \Gamma_{12}^{uc} + \lambda_u^2\,\Gamma_{12}^{uu})\,.
\eeqa
In the OPE treatment they differ only with respect to quark content in loops:  two charm quarks  vs. one charm and one up quark, and satisfy
$\Gamma_{12}^{cc} \cong \Gamma_{12}^{uc}$
\cite{lenznierste}.
This is likely to be the dominant contribution to $\delta \phi_\Gamma $,
certainly if new physics only enters through gluonic or electroweak penguins. 
With sufficient statistics it could be possible to isolate and measure  $\delta \phi_\Gamma $ via Eq. (\ref{eq:SfaSLinBsmod}) applied to $B_s\to J/\Psi \phi$.
This would require that 
the hierarchy in 
Eq. (\ref{eq:gettingphiGamma}), equivalent to $|\delta \phi_\Gamma /2| \gg| \delta\phi_{\lambda_{J/\Psi \phi}}|$, is satisfied.
In practice, a substantial improvement of the direct CP asymmetry bounds for $b \to c \bar c s$ transitions would also be required.

\section{Discussion and Conclusion}\label{sec:conclusion}

If $\phi_{12} ={\rm arg } (M_{12} /\Gamma_{12})$ is the only source of CP violation in neutral meson decays,
then CP violation in pure mixing (CPVMIX), i.e., $|q/p|\ne 1$, and CP violation in the interference of decays with and without mixing (CPVINT), i.e., $\phi \ne 0$,
are related phenomena.   Moreover, $\phi$ would be related to the underlying 
mixing parameters $|M_{12} |$, $|\Gamma_{12} |$, and $\phi_{12}$ of relevance to model building.
New weak phases in the decay amplitudes would enter and modify these relations.
However, existing direct CP violation measurements provide stringent constraints on their magnitudes in the (tree-level dominated) $D^0$ and $B_s$ 
decays of interest to us,
implying that any modifications to the relations between CPVMIX and CPVINT must be small perturbations.  We summarize these results, and their implications 
below.


The general relation between $\phi$ and $|q/p| $ (CPVINT and CPVMIX) in the limit of no new weak phases in decay
is derived in Section IV, see 
Eq.~(\ref{eq:modindepreln}).
It leads to correlations between the semileptonic and time-dependent CP asymmetries and additionally,
in the $D^0$ system, between the semileptonic and time-integrated CP asymmetries.
We remind the reader that in $D^0$ decays the time-dependent ($\Delta Y_f$) and time-integrated ($a_f $)
CP asymmetries must be equal
in the limit of no direct CP violation \cite{Grossman:2006jg} (no new weak phases in decay), see Section III.

Below we will refer to the whole complex of relations obtained via applications of Eq.~(\ref{eq:modindepreln})
as the CPVMIX/CPVINT relations.
We give them a fairly complete treatment in the case of tree-level dominated $D^0$ decays, covering
singly Cabibbo suppressed (SCS) decays to CP ($K^+ K^-$, $\pi^+ \pi^-$) and non-CP ($K^* K$) eigenstates,
as well as Cabibbo favored (CF) and doubly Cabibbo suppressed (DCS) decays to CP eigenstates ($K_s \pi^0$) and to ``wrong-sign" non-CP eigenstates ($K^+ \pi^-$), where examples are included in parentheses.
In the case of $B_s $ decays, we confirm the correlation between the semileptonic and time-dependent  
CP asymmetries obtained in \cite{Grossman:2009mn} for decays to CP eigenstates, and we also obtain the correlation for decays to non-CP eigenstates,
see Eqs.~(\ref{eq:SfaSLinBs}) and (\ref{eq:SfaSLnonCP}).

For SCS $D^0$ decays to non-CP eigenstates, CP conjugate decay rates are of same order.  Therefore,
pairs of time-dependent and time-integrated CP asymmetries are accessible to experiment.
We find that the relation between $\phi$ and $|q/p| $ implies that the ratio of CP asymmetries within each pair is given by the inverse ratio of the corresponding CP averaged decay rates, see Eq.~(\ref{eq:DeltaYBRreln}).
By contrast, the near equality of CP conjugate pairs of time-dependent CP asymmetries ($S_f$ and $S_{\overline f}$) for $B_s$ decays to non-CP eigenstates
is a trivial consequence of $y\ll x$ and $\left| |q/p|-1 \right| \ll1$.

The general expression derived for $\phi$ in terms of  the underlying mixing parameters 
$|M_{12} |$, $|\Gamma_{12} |$, and $\phi_{12}$, in the limit of no new weak phases in decay, is given in Eq. (\ref{eq:fiteqphi}).
It can be combined with similar expressions for $x$, $y$, and $|q/p |$, see Eqs.~(\ref{eq:xfiteq}),~(\ref{eq:yfiteq}), and (\ref{eq:qovp4}),
to extract the underlying $D^0 - \overline{D^0}$ mixing parameters $|M^D_{12} |$, $|\Gamma^D_{12} |$, and $\phi^D_{12}$
from a direct fit to the experimental data.  In this work we adopt the simpler strategy of extracting the mixing parameters from a fit to the HFAG outputs, which include $x$, $y$, $|q/p |$, and $\phi$.  The (parabolic) HFAG output error matrix is used to construct a $\chi^2$
function, see Section V. 
We find that 
\begin{itemize}
\item $\phi^D_{12}$ is currently being probed at the level of $0.10$ [rad] at $1\sigma$, see Eq. (\ref{eq:phi12bnd}).
\item Incorporating the relation between $\phi$ and $\phi_{12}$, Eq. (\ref{eq:fiteqphi}), into the fit 
reduces the experimental errors on $\phi_{12}^D$ by a factor of three for the current data set.
\end{itemize}
The preliminary (non-parabolic) HFAG fit directly to the data \cite{alantalk}, also obtained using Eqs.~(\ref{eq:xfiteq}),~(\ref{eq:yfiteq}), (\ref{eq:qovp4}), and
(\ref{eq:fiteqphi}) for $x$, $y$, $|q/p |$, and $\phi$, 
yields a sensitivity to $\phi_{12}^D$ of 0.10 [rad] ($1\sigma$); 
the 95\% CL bound is $\phi_{12}^D \le 0.30$ [rad], see Eq. (\ref{eq:HFAG1sigma}).


Two questions concerning the impact of new weak phases in decay need to be addressed:
(i) to what extent can the CPVMIX/CPVINT relations be violated
in the $D^0$ and $B_s$ systems,
and how well could such violations be measured in the future;
(ii) to what extent can the bound on $\phi_{12}^D$ be modified.
The violations in (i) can be characterized precisely in terms of the shifts $\delta \phi_{\lambda_f} $
in the CPVINT observables $\phi_{\lambda_f} = {\rm arg}|q \overline{A}_f / p A_f |$ with respect to $\phi$, the shift $\delta \phi_\Gamma $ in $ {\rm arg}(\Gamma_{12} /\Gamma_{12}^* )$ with respect to 
the appropriate leading SM contribution, and the direct CP asymmetries $a_f^d$ (see Eqs.~(\ref{eq:shiftinYf})-- (\ref{eq:DeltaYfnonCPrelnmods2}), Eqs.~(\ref{eq:definitions2})--(\ref{eq:phipphim}), and Eq.~(\ref{eq:SfaSLinBsmod}) in Section VI).  Thus, we need to know how large these quantities can be.

Direct CP violation bounds provide stringent constraints on subleading amplitudes containing new weak phases.
Strong phase differences enter as $\cos\delta$ in the direct CP asymmetries, and as $\sin\delta$ in $\delta \phi_{\lambda_f} $  and $\delta \phi_\Gamma $. 
However, we argue that in all cases of interest the new physics amplitudes would have significant strong phase differences 
with respect to the leading SM amplitudes (due, essentially, to different color and chirality structures for the underlying effective operators).
Therefore, the direct and time-integrated CP violation bounds also translate into order of magnitude bounds on $\delta \phi_{\lambda_f} $, $\delta \phi_{\lambda_{\overline f}} $, and $\delta \phi_\Gamma$
due to new weak phases in decay
(see Eqs.~(\ref{eq:phibndsSCS}), (\ref{eq:phibndsCFDCS}), and (\ref{eq:phibndsBs}) for the SCS, CF/DCS, and 
$b\to c\bar c s$ transitions, respectively).

The main implications of these bounds for $D^0 - \overline{D^0}$ and $B_s -\overline{B_s}$ mixing today are:
\vspace{0.05cm}
\begin{itemize}
\item In SCS $D^0$ decays the maximal allowed violations of the CPVMIX/CPVINT relations are of $O({\rm a~few\,\%})$ of the current
experimental errors on the time-dependent CP asymmetries (CPVINT), see Eqs.~(\ref{eq:shiftinYf})--(\ref{eq:DeltaYfnonCPrelnmods1}).
\item In CF/DCS $D^0$ decays the maximal allowed violations of the CPVMIX/CPVINT relations are of $O(10\,\%)$ of the current
experimental errors on the time-dependent CP asymmetries (CPVINT), see Eqs.~(\ref{eq:definitions2})--(\ref{eq:phiKsPiPi}).   
\item Violations of Eq. (\ref{eq:fiteqphi}), relating $\phi$ and $\phi^D_{12}$, are similarly bounded relative to  
the present day experimental sensitivity to $\phi$ in SCS and CF/DCS $D^0$ decay modes, respectively.
\item Consequently, the bounds on $\phi_{12}^D $ in Eqs.~(\ref{eq:phi12bnd}) and (\ref{eq:HFAG1sigma}) can not be significantly modified by
new weak phases in decay, see Eq. (\ref{eq:phi12fitmodified}).
\item For $b \to c \bar c s$ transitions, the maximal allowed violation of the $B_s$ CPVMIX/CPVINT relations is 
$O(5\%)$ in absolute terms, see Eq. (\ref{eq:SfaSLinBsmod}).
\end{itemize}

At a high luminosity flavor factory (with 75$\,{\rm ab}^{-1}$), we assume that there will be an order of magnitude improvement in precision for individual $D^0 - \overline{D^0}$ mixing measurements, and in the global fit to the data (a reduction of $\approx 6$ in the errors on $\phi$ and $|q/p| $ for the global HFAG fit is projected in \cite{alancipancomp}).
The error on the $B_s$ semileptonic CP asymmetry at LHCb (with 2$\,{\rm fb}^{-1}$) is expected to be $\delta a_{\rm SL} \approx 0.002$ \cite{muheimLHCb}.
Therefore, our conclusions on the sensitivity of mixing measurements at these facilities to new weak phases in decay are: 
\begin{itemize}
\item Violations of the $D^0 - \overline{D^0}$ CPVMIX/CPVINT relations will be probed at 
the same order as the currently allowed maximal violations (obtained from direct and time-integrated CP violation measurements),
implying that they 
could be difficult to observe at a high luminosity flavor factory.
\item  The ``goodness" of a global fit to the  $D^0 - \overline{D^0}$ mixing data which assumes no new weak phases in decay
would probably be more sensitive to their effects than 
violations of the CPVMIX/CPVINT relations in individual decay modes.
\vspace{0.2cm}
\item The expected error in the $B_s$ semileptonic CP asymmetry at LHCb is prohibitively large
for a meaningful probe of the $B_s$ CPVMIX/CPVINT relations to be carried out.
\vspace{0.10cm}
\end{itemize}
\beqa
\nonumber\eeqa
\vspace{-1.7cm}

In principle, with sufficient statistics it would be possible to determine $\delta \phi_\Gamma $ in the $D^0$ and $B_s$ systems,
if the violations of the CPVMIX/CPVINT relations are much larger than the direct CP asymmetries in
the SCS or CF/DCS transitions and the $b\to c\bar c s$ transitions, respectively (see the discussions at the ends of Sections
VIB and VIC).

We emphasize that the $D^0$ and $D^\pm$ direct CP asymmetry measurements
provide much more sensitive probes 
of new weak phases in decay than the time-dependent CP asymmetries (which correspond to differences between the $D^0$ and $\overline{D^0}$
time of decay profiles).  
Recall that in $D^0$ decays, the direct CP asymmetries are obtained from comparison of the time-integrated and time-dependent CP asymmetries.
The effects of new weak phases in the time-dependent CP asymmetries are necessarily suppressed
by mixing ($x$ or $y$).
Therefore, the most likely scenario at a high luminosity flavor factory is that 
improved precision in the time-integrated or direct CP violation measurements
will imply that the effects of new weak phases in decay
lie beyond the reach of the time-dependent CP asymmetry measurements.
It will of course still be possible to probe for new weak phases in $b \to c\bar c s$ transitions at a super-B factory ($B$ decays) and
at the LHC ($B$ and $B_s$ decays) via direct CP violation measurements.

Finally, and of immediate interest for CP violation in $D^0 -\overline{D^0}$ mixing, 
the bounds on $\phi_{12}^D$ imply that  ${\rm Im} (M_{12}^{\rm NP}) /|M_{12} | $ is being probed at the $10\%$ level at $1\sigma$
(for the usual phase convention in which $M^{\rm SM}_{12}$ is real, and where $M_{12}^{\rm NP}$ is the new physics contribution).
The preliminary HFAG 95\% CL interval for $\phi_{12}^D $ implies that $|{\rm Im} (M_{12}^{\rm NP}) /M_{12} | \le 0.30 $ at 95\% CL.
These results apply to models without new weak phases in decay, or with new weak phases in decay (up to an additive correction
of less than a few percent), see Eqs.~(\ref{eq:sinphi12M})--(\ref{eq:ImM12bnd95CL}), and
Eqs. (\ref{eq:sinphi12modified}), (\ref{eq:ImM12bnd1sigmamodified}).
Examples in which CP violation in $D^0 - \overline{D^0}$ mixing at such levels is possible have recently been discussed in the context of 
supersymmetry, little Higgs models, warped extra dimension models, and the minimal flavor violation framework \cite{Blum:2009sk,Bigi:2009df,Gedalia:2009kh,Csaki:2009wc,Kagan:2009bn}.

{\bf Acknowledgements.} 
We thank Yuval Grossman and especially Yossi Nir for numerous discussions on the mixing and CP violation formalism,
Gilad Perez for a discussion of the bound on CP violation in $D^0 - \overline{D^0}$ mixing,
and Alan Schwartz for providing us with the HFAG output correlations (for $A_D = 0$), the preliminary HFAG fit results for $\phi_{12}$, and 
discussions.
A.~L.~K. is supported by DOE grant FG02-84-ER40153,
M.~D.~S. by NSF grant PHY0757876.

\end{document}